\documentclass[aps,a4paper,nofootinbib,showpacs,preprintnumbers,amsmath,amssymb,reqno]{amsart}

\usepackage[utf8]{inputenc}
\usepackage{amsmath,amsthm,amsfonts,latexsym,amssymb,bm,enumerate}
\usepackage{ae}
\usepackage{cite}
\usepackage{float}
\usepackage{lmodern}
\usepackage[T1]{fontenc}
\usepackage{psfrag}

\usepackage{amsmath}
\usepackage{amsthm,amssymb,amstext,amscd,amsfonts,amsbsy,amsxtra,latexsym}

\usepackage{graphics}
\usepackage{graphicx}
\usepackage{color}
\usepackage{hyperref}

\definecolor{dark-red}{rgb}{.54,.0,.0}
\definecolor{dark-green}{rgb}{.0,.4,.0}
\definecolor{dark-blue}{rgb}{.04,.04,.4}

\DeclareFontFamily{OT1}{pzc}{}
\DeclareFontShape{OT1}{pzc}{m}{it}%
{<-> s * pzcmi8t}{}
\DeclareMathAlphabet{\mathpzc}{OT1}{pzc}%
{m}{it}

\hypersetup{colorlinks=true, linkcolor=dark-red, urlcolor=dark-blue, citecolor=dark-green, linktoc=page}

\usepackage[english]{babel}

\newtheorem{thm}{Theorem}[section]

\newtheorem{rmk}[thm]{Remark}

\renewcommand{\epsilon}{\varepsilon}
\renewcommand{\leq}{\leqslant}
\renewcommand{\geq}{\geqslant}

\begin{document}
	
\title[The effect of a $\delta$ distribution potential]{The effect of a $\delta$ distribution potential\\ on a quantum mechanical particle in a box}
  
\author{Pedro Martins Girão}
\author{João Pedro Nunes}
 
\address{Pedro Martins Gir\~ao: 
	Center for Mathematical Analysis, Geometry and Dynamical Systems,
	Instituto Superior T\'ecnico, Universidade de Lisboa, Portugal}
\email{pgirao@math.tecnico.ulisboa.pt}

\address{Jo\~{a}o Pedro Nunes: 
	Instituto Superior T\'ecnico, Universidade de Lisboa, Portugal
}
\email{joaopedrocorreianunes@tecnico.ulisboa.pt}


\thanks{Pedro M.\ Girão was partially funded by FCT/Portugal through project UIDB/04459/2020.}

    \begin{abstract} We study the effect of a  $\delta$ distribution potential
    	placed at $x_0\geq 0$ and multiplied by a parameter $\alpha$ on a quantum mechanical particle in an infinite square well
    	over the segment $\left[-\,\frac{L}{2},\frac{L}{2}\right]$.
    	We obtain the limit of the eigenfunctions of the time independent
    	Schr\"{o}dinger equation as $\alpha\nearrow+\infty$ and as 
    	$\alpha\searrow-\infty$. We see how each solution of the
    	Schr\"{o}dinger equation corresponding to $\alpha=0$ changes as
    	$\alpha$ runs through the real line.
    	When $x_0$ 
is a rational multiple of $L$,
there exist solutions of the Schr\"{o}dinger equation 
which vanish at $x_0$ and are unaffected by the value of~$\alpha$.
We show that each one of these has an energy that coincides with the energy
of a certain limiting eigenfunction obtained by taking $|\alpha|\to\infty$.
The expectation value of
the position of a particle with wave function equal to the limiting eigenfunction
is~$x_0$.
    \end{abstract}
    
 \maketitle
    
\section{Introduction}

This work was motivated by~\cite{jviana11}, where the 
authors study the effect of a $\delta$ distribution potential
placed at the origin on the quantum harmonic oscillator. 
The authors show that, 
up to a sign change in half the domain,
as the constant $\alpha$ multiplying the $\delta$ distribution potential
goes to infinity, each even parity solution converges to the adjacent odd parity solution
(which is not affected by the $\delta$ distribution potential), with a higher angular wave number
if the constant $\alpha\nearrow+\infty$, and with a lower angular wave number if
the constant $\alpha\searrow-\infty$. 
We wanted to understand why this occurs. Additionally, 
the question arose of what would happen if the $\delta$ distribution potential
was not placed at the origin. It seemed clear that new phenomenon 
would have to arise since 
the symmetry of the problem would be broken, and
there might not be any eigenfunctions of the Schr\"{o}dinger operator
that would be unaffected by the $\delta$ distribution potential.
Since it is somewhat cumbersome to work with hypergeometric functions, we decided to consider simplest possible setting, arguably
an infinite square well with a $\delta$ distribution potential.
This paper contains the results we encountered. 

The modes of a quantum particle of mass $m>0$ in a one
dimensional box $\left[-\frac{L}{2},\frac{L}{2}\right]$ ($L>0$),
with a $\delta$
distribution potential placed at $x_0\in\left[0,\frac{L}{2}\right)$ are described 
by the time independent Schr\"{o}dinger equation
\begin{equation}\label{Sch}
\begin{cases}
	-\frac{\hbar^{2}}{2 m} \frac{d^2\Psi}{dx^2} + \alpha \delta(x - x_0) \Psi = E \Psi,\quad \text{for}\ x\in\left(-\,\frac{L}{2},\frac{L}{2}\right), \\
	\Psi\left(-\,\frac{L}{2}\right)= \Psi\left(\frac{L}{2}\right) = 0.
\end{cases}  
\end{equation}
Define 
$$
\underline{\mathcal{P}}:=\left\{\frac{2k\pi}{\frac{L}{2}+x_0}: k\in\mathbb{N}\right\}, \quad
\overline{\mathcal{P}}:=\left\{\frac{2l\pi}{\frac{L}{2}-x_0}: l\in\mathbb{N}\right\},
$$
$$
\mathcal{P}:=\underline{\mathcal{P}}
\cup \overline{\mathcal{P}}\quad \text{and}\quad
\mathcal{K}:=\underline{\mathcal{P}}
\cap \overline{\mathcal{P}}.
$$
If $\underline{\nu}_k\in\underline{\mathcal{P}}$, then  $\underline{\nu}_k$ is
the angular wave number
corresponding to eigenfunctions on
$\left[-\frac{L}{2},x_0\right]$ with $k$ periods, and if
$\overline{\nu}_l\in\overline{\mathcal{P}}$, then $\overline{\nu}_l$ is
the angular wave number corresponding to eigenfunctions on $\left[x_0,\frac{L}{2}\right]$
with $l$ periods. If $\alpha=0$, then~\eqref{Sch} has solutions
$$
\Phi_{{\nu}}(x)=\sqrt{\frac{2}{L}}\sin
\left(\frac{{\nu}}{2}\left(\frac{L}{2}-x\right)\right),\qquad
E_\nu=\frac{\hbar^2}{2m}\left(\frac{\nu}{2}\right)^2,
$$
for
$$
\nu=\nu_n=\frac{2n\pi}{L},\quad n\in\mathbb{N},
$$
$\nu_n$ the angular wave number of a function with $n$ periods in an interval of length~$L$.
We will say that $\Phi_{\nu_n}$ has $n$ oscillations in
$\left[-\,\frac{L}{2},\frac{L}{2}\right]$, an oscillation occurring in an
interval with length equal to half a period of a sine function.

The function $\Phi_{\nu_n}$ vanishes at $x_0$ if and only if
$\nu_n$ belongs to $\mathcal{K}$. In this situation, $\Phi_{\nu_n}$ is a
solution of~\eqref{Sch} for all values of $\alpha$. Suppose now that
$\nu_n$ does not belong to $\mathcal{K}$ and, starting from $\alpha=0$, we
let $\alpha$ run through the real line. We want to understand how 
the solution to~\eqref{Sch} evolves. Particularly, we would like to 
obtain the limit solutions when $\alpha\nearrow+\infty$ and 
$\alpha\searrow-\infty$.

The set $\mathcal{P}$ divides the real line into a family of open subintervals.
If $\nu_n\not\in\mathcal{K}$, then~$\nu_n$ belongs to one of these subintervals,
$\mathcal{I}$.
It turns out that as $\alpha$ increases from~$-\infty$ to~$+\infty$
along the real line,
the parameter $\nu$ increases across $\mathcal{I}$ according to
\begin{equation}\label{x0nu}
	\alpha=\alpha_\nu=\begin{cases}
		-\,\frac{\hbar^2}{2m}\frac{\nu}{2}\frac{\sin\left({ \frac{\nu}{2}L}\right)}{\sin\left({\frac{\nu}{2}\left(\frac{L}{2} + x_0\right)}\right) \sin\left({\frac{\nu}{2}\left(\frac{L}{2} - x_0\right)}\right)}
		&\text{for}\ \nu> 0,\\
		-\,\frac{\hbar^2}{2m}\frac{4L}{L^2-4x_0^2}&\text{for}\ \nu=0,\\
		-\,\frac{\hbar^2}{2m}\frac{\nu}{2}\frac{\sinh\left({ \frac{\nu}{2}L}\right)}{\sinh\left({\frac{\nu}{2}\left(\frac{L}{2} + x_0\right)}\right) \sinh\left({\frac{\nu}{2}\left(\frac{L}{2} - x_0\right)}\right)}&\text{for}\ \nu<0.
	\end{cases}
\end{equation}
The function~\eqref{x0nu} is a diffeomorphism from $\mathcal{I}$ to~$\mathbb{R}$, 
except when $\mathcal{I}=(-\infty,\underline{\nu}_1)$. In this last situation,
\eqref{x0nu} is an homeomorphism from $(-\infty,\underline{\nu}_1)$ to $\mathbb{R}$,
with a positive derivative at every point which is not~$0$.
The Schr\"{o}dinger equation~\eqref{Sch},
with $\alpha=\alpha_\nu$ and
$$
E=E_\nu=\begin{cases}
	\frac{\hbar^2}{2m}\left(\frac{\nu}{2}\right)^2&\text{for}\ \nu\geq 0,\\
	-\,\frac{\hbar^2}{2m}\left(\frac{\nu}{2}\right)^2&\text{for}\ \nu< 0,
\end{cases}
$$
 has a weak solution
$\Psi_\nu$, 
given in~\eqref{psi} for $\nu>0$, in~\eqref{psizero} for $\nu=0$, and
in~\eqref{psinegativo} for $\nu<0$. We call
$\rho_\nu$ the norm of $\Psi_\nu$. 

We define $$\mathpzc{q}:=
\frac{\frac{L}{2}-x_0}{\frac{L}{2}+x_0}.$$
Our main result is
\begin{thm}
	When an endpoint $\underline{\nu}$ of $\mathcal{I}$ belongs to $\underline{\mathcal{P}}\setminus
	\overline{\mathcal{P}}$, then 
	$\nu$ goes to $\underline{\nu}$ as $\alpha\searrow-\infty$ if $\underline{\nu}$ is the lower endpoint, and $\nu$ goes to $\underline{\nu}$ as $\alpha\nearrow+\infty$ if $\underline{\nu}$ is the upper endpoint.
	The limiting solution\/
	$
	\lim_{\nu\to\underline{\nu}}\frac{\Psi_\nu}{\rho_\nu}
	$,
	$\underline{\Upsilon}_{\underline{\nu}}$,
	is zero on $\left[x_0,\frac{L}{2}\right]$.
	The set $\underline{\mathcal{P}}\setminus\overline{\mathcal{P}}$ is empty
	when $x_0=0$.
	
	When an endpoint $\overline{\nu}$ of $\mathcal{I}$ belongs to $\overline{\mathcal{P}}\setminus
	\underline{\mathcal{P}}$, then 
	$\nu$ goes to $\overline{\nu}$ as $\alpha\searrow-\infty$ if $\overline{\nu}$ is the lower endpoint, and $\nu$ goes to $\overline{\nu}$ as $\alpha\nearrow+\infty$ if $\overline{\nu}$ is the upper endpoint.
	The limiting solution\/
		$
	\lim_{\nu\to\overline{\nu}}\frac{\Psi_\nu}{\rho_\nu}
	$,
	$\overline{\Upsilon}_{\overline{\nu}}$,
	is zero on $\left[-\,\frac{L}{2},x_0\right]$. 
	The set $\overline{\mathcal{P}}\setminus\underline{\mathcal{P}}$ is empty
	when $\mathpzc{q}=\frac{1}{m}$, for some natural $m$.
	
	When an endpoint $\hat{\nu}$ of $\mathcal{I}$ belongs to $\mathcal{K}$,
	we have\/ $\lim_{\nu\to\hat{\nu}}\rho_\nu=0$ and
	$$
	\Upsilon_{\hat{\nu}}(x):=
	\lim_{\nu\to\hat{\nu}}\frac{\Psi_\nu(x)}{\rho_{\nu}}=
	\begin{cases}
		-\sqrt{\mathpzc{q}}\,
		\Phi_{\hat{\nu}}(x), &
		\text{for}\ -\frac{L}{2}\leq x\leq x_0,\\
		\frac{1}{\sqrt{\mathpzc{q}}}\,
		\Phi_{\hat{\nu}}(x), & \text{for}\ x_0<x\leq \frac{L}{2}.
	\end{cases}
	$$
	The convergence is in $C^0\left[-\,\frac{L}{2},\frac{L}{2}\right]$,
	in $C^1\left[-\,\frac{L}{2},x_0\right]$, and in
	$C^1\left[x_0,\frac{L}{2}\right]$.
	$\Upsilon_{\hat{\nu}}$ is a weak solution of the limit equation~\eqref{eq1_alt}.
	We have
	\begin{equation}\label{chico}
	\frac{
		\int_{x_0}^{\frac{L}{2}}\Upsilon_{\hat{\nu}}^2(x)\,dx}
	{\int_{-\frac{L}{2}}^{x_0}\Upsilon_{\hat{\nu}}^2(x)\,dx}
	=\frac{1}{\mathpzc{q}}>1,
\end{equation}
	except when $x_0=0$, in which case $\mathpzc{q}=1$.
	In addition, 
	\begin{equation}\label{lira}
	\int_{-\,\frac{L}{2}}^{\frac{L}{2}}x\Upsilon_{\hat{\nu}}^2(x)\,dx=x_0.
\end{equation}
	The Fourier series of $\Upsilon_{\hat{\nu}}$ is given
	in~\eqref{Fourier}. In particular, 
	$\Upsilon_{\hat{\nu}}$ is $L^2$ orthogonal to 
	the modes which vanish at $x_0$.
	The set $\mathcal{K}$ is empty when $x_0$ is an irrational multiple of~$L$.
		
	 Finally, we have 
	$$
	\lim_{\nu \to -\infty}\frac{
		\int_{x_0}^{\frac{L}{2}}\Psi_{\nu}^2(x)\,dx}
	{\int_{-\frac{L}{2}}^{x_0}\Psi_{\nu}^2(x)\,dx}=1,
	$$
	$$
	\lim_{\nu \to -\infty}
	\int_{-\,\frac{L}{2}}^{\frac{L}{2}}x\left(\frac{\Psi_\nu}{\rho_\nu}\right)^2(x)\,dx=x_0.
	$$
\end{thm}

If we start with $\alpha=0$, at the ground state, $\Phi_{\nu_1}$ and decrease $\alpha$,
 below a large negative number,
then the particle will tend to be close to $x_0$, where the potential is
very negative, with almost equal probability of being on either side of $x_0$.
If $x_0>0$ and we increase $\alpha$ above a large positive number,
then the particle will tend to be on the left of $x_0$, 
as the interval $\left(-\,\frac{L}{2},x_0\right)$ has a bigger length
 than the interval $\left(x_0,\frac{L}{2}\right)$,
and thus accommodates a wave with a lower angular wave number.

For states which are not the ground state, having a large negative
Dirac $\delta$ potential does not raise the probability of finding the 
particle around $x_0$. In fact, the opposite is true. 
No matter what the sign 
of $\alpha$, as long as the absolute value of $\alpha$ is large,
then the effect of a large Dirac $\delta$ potential,
on a state, which is not a ground level state, is to lower the probability of finding the particle in a very small neighborhood of~$x_0$, forcing the wave function to vanish at $x_0$, in the limit as~$|\alpha|\to \infty$.

\begin{figure}[ht!]
	\begin{psfrags}
		\psfrag{n}{$\nu$}
		\psfrag{f}{$\!\!\!\!\!\!\!\!\Phi_{\nu_{16}}(x)$}
		\psfrag{u}{$\!\!\!\!\!\!\!\!\Upsilon_{\nu_{16}}(x)$}
		\psfrag{b}{{\tiny $x_0$}}
		\psfrag{c}{$x$}
		\psfrag{d}{$\frac{L}{2}$}
		\psfrag{z}{$0$}
		\psfrag{zz}{\!\,$0$}
		\psfrag{1}{$\!\!\!\!\nu_1$}
		\psfrag{2}{$\!\!\!\!\nu_2$}
		\psfrag{3}{$\!\!\!\!\nu_3$}
		\psfrag{4}{$\!\!\!\!\nu_4$}
		\psfrag{5}{$\!\!\!\!\nu_5$}
		\psfrag{6}{$\!\!\!\!\nu_6$}
		\psfrag{7}{$\!\!\!\!\nu_7$}
		\psfrag{8}{$\!\!\!\!\nu_8$}
		\psfrag{9}{$\!\!\!\!\nu_9$}
		\psfrag{10}{\!\!\!\!$\nu_{10}$}
		\centering
		\includegraphics[scale=1.1]{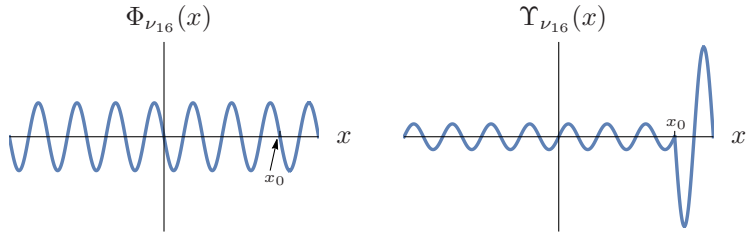}
		\caption{Sketch of the graphs of $\Phi_{\nu_{16}}$ and
			$\Upsilon_{\nu_{16}}$, in the case that $x_0=\frac{3L}{8}$.
			Here $\nu_{16}=16\,\frac{2\pi}{L}=
			14\,\frac{2\pi}{\frac{L}{2}+x_0}=2\,\frac{2\pi}{\frac{L}{2}-x_0}$.
			The value of $\mathpzc{q}$ is equal to~$\frac{1}{7}$.
			Both plots share an identical vertical axis scale.
		}
		\label{exemplo}
	\end{psfrags}
\end{figure}

If $\hat{\nu}$ belongs to $\mathcal{K}$, then $\hat{\nu}=\underline{\nu}_k\in
\underline{\mathcal{P}}$, and  
$\hat{\nu}=\overline{\nu}_l\in\overline{\mathcal{P}}$, for some $k$ and $l$.
The value $\frac{\hat{\nu}}{2}$ is the angular wave number
corresponding to eigenfunctions on the interval
$\left[-\frac{L}{2},x_0\right]$ and
corresponding to eigenfunctions on the interval $\left[x_0,\frac{L}{2}\right]$.
We then have $\hat{\nu}=\nu_{k+l}$ and $\Phi_{\nu_{k+l}}(x_0)=0$;
$\Phi_{\nu_{k+l}}$ has~$k$
oscillations on the left of $x_0$, and~$l$ oscillations on the right of $x_0$.
For a particle described by $\Phi_{\hat{\nu}}$, the ratio of the probability
of finding the particle on the right of $x_0$ over the
probability of finding the particle on the left of $x_0$, is equal to the ratio of the lengths of these two subintervals,
which is $\mathpzc{q}$. As we have observed above, this wave function is unaffected by the value of $\alpha$.  For
a particle described by the limiting wave function
$\Upsilon_{\hat{\nu}}$, which has the same energy,
 the ratio of the probability
on the right of $x_0$ over the
probability on the left of $x_0$ is equal to~$\frac{1}{\mathpzc{q}}$
(equality~\eqref{chico}).
Using commonly adopted terminology in this context, we may say that the state $\Phi_{\hat{\nu}}$ is quasi-degenerate.
In Figure~\ref{exemplo} we sketch the graphs of $\Phi_{\nu_{16}}$ and
$\Upsilon_{\nu_{16}}$, in the case that $x_0=\frac{3L}{8}$.

Although the probability of finding the particle in a very small neighborhood of~$x_0$
is small, the expectation value of
the position of a particle described by~$\Upsilon_{\hat{\nu}}$ is equal to $x_0$
(equality~\eqref{lira}). 
This is a natural result. It sheds light on why, at an angular wave number which
corresponds to eigenfunctions on 
 $\left[-\,\frac{L}{2},x_0\right]$ 
 and on $\left[x_0,\frac{L}{2}\right]$, 
 according to the limiting wave function $\Upsilon_{\hat{\nu}}$,
the probability of finding a particle on the smaller right subinterval
is bigger than the probability of finding the particle on the larger left
subinterval.

Consider the situation where $\underline{\nu}_k$,
$\underline{\nu}_{k+1}\in\underline{\mathcal{P}}\setminus
\overline{\mathcal{P}}$ and
$(\underline{\nu}_k$, $\underline{\nu}_{k+1})\cap\mathcal{P}=\emptyset$.
Both as~$\nu$
decreases to~$\underline{\nu}_k$,
and as~$\nu$
increases to~$\underline{\nu}_{k+1}$, the
eigenfunctions $\Psi_\nu$
describe particles with a high probability of being on the left of
$x_0$. We show that, as we increase~$\nu$ 
from~$\underline{\nu}_{k}$ to~$\underline{\nu}_{k+1}$, we may find values for which 
$\Psi_\nu$ describes a particle with a high probability of being on the right of~$x_0$
(even though there is no value of $\overline{\mathcal{P}}$ between~$\underline{\nu}_{k}$ and~$\underline{\nu}_{k+1}$, see Subsection~\ref{surprise}).

Let $n$ be odd, and $\nu_n=\frac{2n\pi}{L}$,
so that 
$\Phi_{\nu_n}$
is an even wave
function with $n$ oscillations.
In the case where $x_0=0$,
we study with particular
attention the change in the amplitude of the wave functions
$\frac{\Psi_\nu}{\rho_\nu}$, with~$\nu$,  seeing how
within each interval $(\nu_{n-1},\nu_n)$ they increase and then return to their original value, 
and within each interval $(\nu_{n},\nu_{n+1})$ they decrease and then return to their original
value
(see Remark~\ref{swing}). (Here we consider $\nu_0=0$ if $n=1$.)

\begin{figure}[ht!]
	\begin{psfrags}
		\psfrag{n}{$\nu$}
		\psfrag{x}{$x_0$}
		\psfrag{a}{$\frac{L}{8}$}
		\psfrag{b}{$\frac{L}{4}$}
		\psfrag{c}{$\frac{3L}{8}$}
		\psfrag{d}{$\frac{L}{2}$}
		\psfrag{z}{$0$}
		\psfrag{zz}{\!\,$0$}
		\psfrag{1}{$\!\!\!\!\nu_1$}
		\psfrag{2}{$\!\!\!\!\nu_2$}
		\psfrag{3}{$\!\!\!\!\nu_3$}
		\psfrag{4}{$\!\!\!\!\nu_4$}
		\psfrag{5}{$\!\!\!\!\nu_5$}
		\psfrag{6}{$\!\!\!\!\nu_6$}
		\psfrag{7}{$\!\!\!\!\nu_7$}
		\psfrag{8}{$\!\!\!\!\nu_8$}
		\psfrag{9}{$\!\!\!\!\nu_9$}
		\psfrag{10}{\!\!\!\!$\nu_{10}$}
		\centering
		\includegraphics[scale=1]{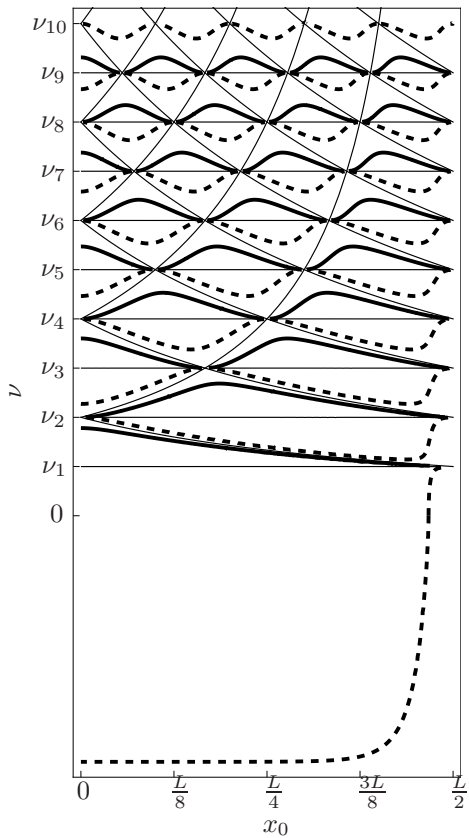}
		\caption{The solutions of~\eqref{x0nu} for a positive~$\alpha$
			(thick solid lines), for $\alpha=0$ (thin horizontal lines),
			and for a negative~$\alpha$ (thick dashed lines).
			The thin solid lines where $\nu$ decreases with $x_0$
			represent $\underline{\mathcal{P}}$, and
			the thin solid lines where $\nu$ increases with $x_0$
			represent $\overline{\mathcal{P}}$.
		}
		\label{x_nu}
	\end{psfrags}
\end{figure}

The problem of a particle in a box with a Dirac~$\delta$ potential
has been previously studied in~\cite{J}. In particular,
the author obtains the expressions for $\Psi_\nu$ in~\eqref{psi}
and in~\eqref{psinegativo}, as well as the relationship between
$\alpha$ and~$\nu$ expressed in~\eqref{x0nu}. He also draws 
Figure~\ref{x_nu} (without the thin lines,
corresponding to $\underline{\mathcal{P}}$ and
$\overline{\mathcal{P}}$), summarizing what happens to~$\nu$ as~$|\alpha|$
goes to~$+\infty$, pointing out the degeneracy in the limit.

The equations~\eqref{x0nu} and~\eqref{psi} were also obtained in~\cite{PV}.
This paper uses the factorization method.


We also mention the work~\cite{GGN}, where the authors 
studied the infinite square well with a singular potential that might not be
placed at the center of the well. Among other findings, they showed that if the singular potential is not centered, then
all energy levels undergo a shift, and they obtain~\eqref{x0nu}.

This paper is organized as follows. 
In Section~\ref{sec2} we set up the problem.
In Sections~\ref{pos}, \ref{zero} and~\ref{neg} we look at positive,
zero and negative energies, respectively.
In Section~\ref{exemplos}, we give some examples illustrating the behavior of the solutions in each of
the subintervals of $\mathbb{R}$ determined by $\mathcal{P}$.
In Section~\ref{pr} we look into the behavior of the ratio of the probability of finding a particle on the right of $x_0$ over the probability of finding a particle
 on the left of $x_0$, as a function of~$\nu$, and we look at the 
expectation value of
the position of a particle, as a function of~$\nu$.
In Appendix~\ref{serie}, we calculate the Fourier series of 
$\frac{\Psi_\nu}{\rho_\nu}$.
In Appendix~\ref{app}, we look at the amplitude of the wave function, 
in terms of $\nu$, in the case that $x_0=0$.

\vspace{3mm}

\noindent {\bf Acknowledgments.} The authors thank Guilherme Milhano for recommending the reading of~\cite{jviana11}. They also thank José Mourão for helpful comments.

\section{Infinite square well with a Dirac $\delta$ potential}\label{sec2}

We refer the reader who is not familiar with
quantum mechanics in one space dimension for a particle
that is confined to move in a ``box''
to~\cite[Section~3.9]{hall13}.
We determinate the spectrum and eigenfunctions of a particle in the interval 
$\left[-\,\frac{L}{2},\frac{L}{2}\right]$ with a Dirac $\delta$ potential at $x_0\in\left[0,\frac{L}{2}\right)$ (the case of negative $x_0$ is immediately 
obtained by considering the reflection $x\mapsto -x$).
We say that $\Psi$ is an eigenfunction, corresponding to the eigenvalue $E$,
if $\Psi\in W^{1,2}\left(-\,\frac{L}{2},\frac{L}{2}\right)$ is nonzero
and is a weak solution of
the time-independent Schr\"odinger equation with homogeneous Dirichlet boundary conditions:
\begin{equation}\label{eq1}
  \begin{cases}
   -\frac{\hbar^{2}}{2 m} \frac{d^2\Psi}{dx^2} + \alpha \delta(x - x_0) \Psi = E \Psi, \\
      \Psi\left(-\,\frac{L}{2}\right)= \Psi\left(\frac{L}{2}\right) = 0
  \end{cases}  
\end{equation}
(see~\cite[(3.43) and~(3.44)]{hall13}).
Here $\delta(x - x_0)$ is the distribution defined by
$$
    \delta(x - x_0)(\chi) = \chi(x_0),
$$
for each continuous function $\chi$ with compact support, and $\alpha$ is a real parameter. Note that the product $\delta(x - x_0) \Psi$ is well-defined,
and the boundary conditions make sense, because,
by the Sobolev Embedding Theorem,
functions in $W^{1,2}\left(-\,\frac{L}{2},\frac{L}{2}\right)$ are H\"{o}lder
continuous in $\left[-\,\frac{L}{2},\frac{L}{2}\right]$.\footnote{Recall that
	a weak solution of~\eqref{eq1} is a function 
	$\Psi\in W^{1,2}\left(-\,\frac{L}{2},\frac{L}{2}\right)$
	that 
	satisfies
$\Psi\left(-\,\frac{L}{2}\right)= \Psi\left(\frac{L}{2}\right) = 0$ and
	$$ 
	-\frac{\hbar^{2}}{2 m} \int_{-\frac{L}{2}}^{\frac{L}{2}}\Psi(x)\frac{d^2\chi}{d{x}^2}(x)\,dx + \alpha \int_{-\frac{L}{2}}^{\frac{L}{2}}\delta(x - x_0) \Psi(x)\chi(x)\,dx = E\int_{-\frac{L}{2}}^{\frac{L}{2}} \Psi(x)\chi(x)\,dx,
    $$ 
	for all $\chi\in C^2_0\left(-\,\frac{L}{2},\frac{L}{2}\right)$.
	}
\par
According to an elementary bootstrap argument, 
the restriction of
each solution of~\eqref{eq1} to
 both the intervals 
$\left[-\,\frac{L}{2},x_0\right]$ and $\left[x_0,\frac{L}{2}\right]$ is
smooth (in particular~$C^2$),
as the Schr\"odinger equation restricted to the intervals 
$\left(-\,\frac{L}{2},x_0\right)$ and $\left(x_0,\frac{L}{2}\right)$
is simply 
\begin{equation}\label{simple}
	-\frac{\hbar^{2}}{2 m} \frac{d^2\Psi}{d{x}^2} = E \Psi.
\end{equation}
Integrating~\eqref{eq1} around $x_0$, we find that
\begin{equation}\label{eq4}
    \Psi'(x_0^+) - \Psi'(x_0^-) = \frac{2 m \alpha}{\hbar^2}\Psi(x_0).
\end{equation}
Therefore, the derivative of the wave function $\Psi$ is discontinuous at $x =x_0$,
except in the case that $\alpha=0$. 
We illustrate the shape of the graph of 
$\Psi$ at $x_0$ in the Table~1, according to
the signs of $\Psi(x_0)$ and $\alpha$.
\begin{table}[ht!]
	\centering
	\begin{tabular}{|c|c|c|}
		\hline
		& $\alpha>0$ & $\alpha<0$  \\
		\hline
		$\Psi(x_0)>0$	& $\vee$ & $\wedge$    \\
		\hline
		$\Psi(x_0)<0$	& $\wedge$ & $\vee$ \\
		\hline
	\end{tabular}
	
	\vspace{2mm}
	
	\label{tabua}
	\caption{The shape of the graph of $\Psi$ at $x_0$.}
\end{table}
The weak solutions of~\eqref{eq1} are
the continuous functions in $\left[-\,\frac{L}{2},\frac{L}{2}\right]$
which are classical solutions of~\eqref{simple} in both $\left[-\,\frac{L}{2},x_0\right]$ and $\left[x_0,\frac{L}{2}\right]$,
satisfy homogeneous Dirichlet boundary conditions, and satisfy~\eqref{eq4}.
Indeed, one readily checks that, conversely, if $\Psi$ satisfies these conditions,
then clearly $\Psi'\in L^2\left(-\,\frac{L}{2},\frac{L}{2}\right)$, and
$\Psi$ is a weak solution of~\eqref{eq1}.

Multiplying both sides of the Schr\"{o}diger equation by $\overline{\Psi}$
and integrating over the interval $\left[-\,\frac{L}{2},\frac{L}{2}\right]$,
and choosing $\Psi$ with $L^2$ norm equal to $1$, we find that
$$
E=\frac{\hbar^2}{2m}\int_{-\,\frac{L}{2}}^{\frac{L}{2}}\left|\frac{d\Psi}{dx}(x)\right|^2\,dx+\alpha|\Psi(x_0)|^2.
$$
Since we are taking the parameter $\alpha$ to be real,
this shows that $E$ is real. In the next three sections we consider
successively the case where $E$ is positive, zero and negative.

\section{Positive eigenvalues} \label{pos}
We start by looking for solutions with $E>0$.
The solutions of~\eqref{simple} which vanish at $\pm\frac{L}{2}$ are
\begin{equation}\label{eq6}
    \Psi(x) = \begin{cases}
        A \sin\left(\frac{\nu }{2}\left(\frac{L}{2} + x\right)\right), &
        \text{for}\ -\frac{L}{2}<x<x_0,\\
        B \sin\left(\frac{\nu }{2}\left(\frac{L}{2} - x\right)\right), & \text{for}\ x_0<x<\frac{L}{2},
    \end{cases}
\end{equation}
with
\begin{equation}\label{ener}
	\nu = 2\sqrt{\frac{2mE}{\hbar^2}}.
\end{equation}
These solutions are present in~\cite[p.~3]{J} and~\cite[(3)]{PV}, for example.

\subsection{Case when $\sin\left(\frac{\nu }{2}\left(\frac{L}{2} + x_0\right)\right)=
	\sin\left(\frac{\nu }{2}\left(\frac{L}{2} - x_0\right)\right)=0$} 

In this case $\Psi(x_0)$ must
be zero. Moreover, we must have on the one hand
$\nu=\underline{\nu}_k$ with $$\underline{\nu}_k=\frac{2k\pi}{\frac{L}{2}+x_0},$$
and on the other hand
$\nu=\overline{\nu}_l$ with $$\overline{\nu}_l=\frac{2l\pi}{\frac{L}{2}-x_0}.$$
Equation~\eqref{eq4} shows that 
$\Psi'$ is continuous at $x_0$. As we already know that
the restrictions of $\Psi$ to $\left[-\,\frac{L}{2},x_0\right]$ and to
$\left[x_0,\,\frac{L}{2}\right]$ are $C^1$, it follows that $\Psi$ is 
$C^1$ in $\left[-\,\frac{L}{2},\frac{L}{2}\right]$. 
This implies that $A\cos(k\pi)=-B\cos(l\pi)\Leftrightarrow B=(-1)^{l-k+1}A$.
The equality $\underline{\nu}_k=\overline{\nu}_l$ implies that
$x_0$ is a rational multiple of $L$:
\begin{equation}\label{pq}
	\underline{\nu}_k=\overline{\nu}_l\ \Leftrightarrow\
	\frac{2k\pi}{\frac{L}{2}+x_0}=\frac{2l\pi}{\frac{L}{2}-x_0}\
	\Leftrightarrow\ x_0=\frac{k-l}{k+l}\frac{L}{2}.
\end{equation}
Moreover,
\begin{equation}\label{kmaisl}
	\underline{\nu}_k=\frac{2k\pi}{\frac{L}{2}+x_0}
	=\frac{2k\pi}{\frac{L}{2}+\frac{k-l}{k+l}\frac{L}{2}}
	=\frac{(k+l)2\pi}{L}.
\end{equation}
Hence
$$
\nu=
\underline{\nu}_k=\overline{\nu}_l=\nu_n:=\frac{2n\pi}{L},\ \text{with}\ n=k+l.
$$
The function $\Psi$
has $k$ half cycles to the left of $x_0$, and has $l$ half cycles to the
right of $x_0$. This is in accordance with the middle equality of~\eqref{pq},
which can also be written as
\begin{equation}\label{lk}
	\frac{l}{k}=\frac{\frac{L}{2}-x_0}{\frac{L}{2}+x_0}.
\end{equation}
We define 
\begin{equation}\label{qqq}
	\mathpzc{q}:=\frac{\frac{L}{2}-x_0}{\frac{L}{2}+x_0}.
\end{equation}
Clearly, $x_0\in\left[0,\frac{L}{2}\right)$ is a rational multiple of $L$ if and only if $\mathpzc{q}\in(0,1]$ is
a rational number.
We have shown that in the present situation 
$\Psi$ is in fact a multiple of the function
\begin{eqnarray*}
	\Phi_{\nu_n}(x)&:=&\sqrt{\frac{2}{L}}
	\sin\left(\frac{\nu_n}{2}\left(\frac{L}{2} - x\right)\right)\\
	&=&(-1)^{n+1}\sqrt{\frac{2}{L}}\sin\left(\frac{\nu_n}{2}\left(\frac{L}{2} + x\right)\right),\
	\text{with}\ n=k+l.
\end{eqnarray*}
We have normalized $\Phi_{\nu_n}$ to have $L^2$ norm equal to one, and to 
be positive in a left neighborhood of $\frac{L}{2}$.

\begin{rmk}\label{q}
Suppose $x_0$ is the rational multiple $\frac{p}{q}$ of $\frac{L}{2}$, where $p\in\mathbb{N}_0$ and $q\in\mathbb{N}$, with
$p<q$ $\left(\text{as}\ x_0<\frac{L}{2}\right)$. 
We will call
$\mathcal{K}$
the set of positive real numbers~$\nu$ such that $\nu=\underline{\nu}_k=\overline{\nu}_l$.
The equality~\eqref{pq}
holds if the pair~$(k,l)$ 
belongs to~$\mathbb{N}^2$ and is a real positive constant times
$\left(\frac{p+q}2,\frac{q-p}2\right)$. 
Suppose that~$p$ and~$q$ have no common factors.
If $p$ and $q$ are both even, or if~$p$ and~$q$ are both odd,
then~$k+l$ is a natural multiple of~$q$. 
In this case, $\mathcal{K}$ consists of the natural multiples of $\nu_q$.
If~$p$ and~$q$ have different parity,
then $k+l$ is a natural multiple of~$2q$.
In this case, $\mathcal{K}$ consists of the natural multiples of $\nu_{2q}$.
\end{rmk}

The value of the energy is given by
$$
E=
E_{\underline{\nu}_k}:=\frac{\hbar^2}{2m}\frac{\underline{\nu}_k^2}{4}=E_{\overline{\nu}_l}:=\frac{\hbar^2}{2m}\frac{\overline{\nu}_l^2}{4}.
$$
A function $\Phi_{\nu_n}$ as above,
a sine wave with
$k$ half cycles to the left of $x_0$ and $l$ half cycles to the
right of $x_0$ is a solution of the Schr\"{o}dinger equation~\eqref{eq1}
for any value of $\alpha$ (as $\Phi_{\nu_n}(x_0)=0$).

\subsection{Case when $\sin\left(\frac{\nu }{2}\left(\frac{L}{2}+ x_0\right)\right)\sin\left(\frac{\nu }{2}\left(\frac{L}{2} - x_0\right)\right)=0$
but not both factors are zero} 
Suppose first that $\nu=\underline{\nu}_k\neq\overline{\nu}_l$.
This forces $B$ to be zero. As $\Psi(x_0)=0$, \eqref{eq4} implies that
$\Psi$ is $C^1$. This implies that $A$ is also zero, and so $\Psi\equiv 0$.
The other case is identical.

\subsection{Case when $\sin\left(\frac{\nu }{2}\left(\frac{L}{2}+ x_0\right)\right)\sin\left(\frac{\nu }{2}\left(\frac{L}{2} - x_0\right)\right)\neq 0$}
\label{third}

We define
$$
\mathcal{P}:=\underline{\mathcal{P}}\cup\overline{\mathcal{P}}=
\left\{k\underline{\nu}_1: k\in\mathbb{N}\right\}
\cup\left\{l\overline{\nu}_1: l\in\mathbb{N}\right\}.
$$
In this subsection we consider the case of positive $\nu$ not belonging to 
$\mathcal{P}$.
To impose continuity of the wave function at~$x_0$ we choose
\begin{equation}\label{AB}
	\begin{cases}
		A =  (-1)^{\left\lfloor\frac{\nu}{2}\left(\frac{L}{2}+x_0\right)\frac{1}{\pi}
			\right\rfloor}
		\sin\left(\frac{\nu }{2}\left(\frac{L}{2} - x_0\right)\right),\\
		B =  (-1)^{\left\lfloor\frac{\nu}{2}\left(\frac{L}{2}+x_0\right)\frac{1}{\pi}
			\right\rfloor}\sin\left(\frac{\nu }{2}\left(\frac{L}{2} + x_0\right)\right)
		=\left|\sin\left(\frac{\nu }{2}\left(\frac{L}{2} + x_0\right)\right)\right|
		.
	\end{cases}    
\end{equation}
The choice of the sign of $A$ and $B$ fixes $\Psi$ to be positive in a 
right neighborhood of $\frac{L}{2}$.
We will denote by
$
\Psi_\nu
$
the function~\eqref{eq6}, with $A$ and $B$ given by~\eqref{AB}:
\begin{eqnarray}\label{psi}
&&\Psi_\nu(x)=\\
&&\begin{cases}
	(-1)^{\left\lfloor\frac{\nu}{2}\left(\frac{L}{2}+x_0\right)\frac{1}{\pi}
		\right\rfloor}
	\sin\left(\frac{\nu }{2}\left(\frac{L}{2} - x_0\right)\right) \sin\left(\frac{\nu }{2}\left(\frac{L}{2} + x\right)\right), &\!\!\!\!
	\text{for}\ -\frac{L}{2}<x<x_0,\nonumber\\
	\left|
	\sin\left(\frac{\nu }{2}\left(\frac{L}{2} + x_0\right)\right)\right| \sin\left(\frac{\nu }{2}\left(\frac{L}{2} - x\right)\right), & 
	\!\!\!\!
	\text{for}\ x_0<x<\frac{L}{2}.\nonumber
\end{cases}
\end{eqnarray}
To normalize $\Psi_{\nu}$, we calculate
\begin{eqnarray}
	&&	\int_{-\,\frac{L}{2}}^{\frac{L}{2}}\Psi_\nu^2(x)\,dx\label{norm}\\
	&&=\frac{1}{2} \left(
	\left(\frac L2+x_0\right) \left(1-\frac{\sin\left(\nu\left(\frac L2+x_0 \right)\right)}{\nu\left(\frac L2+ x_0 \right)}\right) \sin ^2
	\left(\frac{\nu}{2}\left(\frac L2- x_0 \right)\right)
	\right.\nonumber\\
	&&\ \ \ \ \ \left.+
	\left(\frac L2- x_0\right) \left(1-\frac{\sin\left(\nu\left(\frac L2- x_0 \right)\right)}{\nu\left(\frac L2- x_0 \right)}\right) \sin ^2
	\left(\frac{\nu}{2}\left(\frac L2+ x_0 \right)\right)
	\right)\nonumber\\
	&&=:\rho_\nu^2,\nonumber
\end{eqnarray}
with $\rho_\nu\geq 0$.
Recall that the function $x\mapsto 1-\,\frac{\sin x}{x}$ is positive 
for $x>0$ and has limit equal to $1$ when $x\to+\infty$.
The function $\frac{\Psi_\nu}{\rho_\nu}$ has $L^2$ norm equal to one.
The value of $\rho_\nu$ is zero if and only if $\nu=\underline{\nu}_k=
\overline{\nu}_l$ and we are now working under the hypotheses that
$\nu\neq\underline{\nu}_k$ and $\nu\neq\overline{\nu}_l$.

Up to this point $\nu$ is free (greater than zero).
However, we still have to satisfy~\eqref{eq4}. Let
$$g = \frac{m\alpha}{\hbar^2}.$$
From~\eqref{eq4}, using the formula for the sine of the sum, we obtain
\begin{eqnarray}
    2g &=& 
    \frac{\Psi'(x_0^+) - \Psi'(x_0^-)}{\Psi(x_0)}\nonumber
    \\ &=&-\frac{\nu}{2}\frac{\sin\left({ \frac{\nu}{2}L}\right)}{\sin\left({\frac{\nu}{2}\left(\frac{L}{2} + x_0\right)}\right) \sin\left({\frac{\nu}{2}\left(\frac{L}{2} - x_0\right)}\right)}\  =:\ f(\nu).\label{eq8}
\end{eqnarray}
This equality was obtained in~\cite[(18)]{GGN}, \cite[(2)]{J} and~\cite[(5) and (21)]{PV}.
We consider $f$ defined in $\mathbb{C}\setminus\left(\mathcal{P}\cup\{0\}
\cup(-\mathcal{P})\right)$, although in this section~$\nu$ only takes
values in $\nu\in\mathbb{R}^+\setminus\mathcal{P}$.
An alternative expression for $f$ is
\begin{eqnarray}
f(\nu)
&=&-\,\frac{\nu}{2}\left(
\cot\left(\frac{\nu}{2}\left(\frac{L}{2}-x_0\right)\right)+
\cot\left(\frac{\nu}{2}\left(\frac{L}{2}+x_0\right)\right)
\right).\label{alternative}
\end{eqnarray}
\begin{rmk}
Given $g$, \eqref{eq8} determines $\nu$,
which in turn determines $E$ (by \eqref{ener}).
Alternatively, and this is the point of view that we will take,
instead of thinking of $g$ (or equivalently $\alpha$) as given, we may consider $\nu$ as the independent
parameter. 
\end{rmk}
\noindent Indeed, provided that~$\nu\not\in\mathcal{P}$,
$\Psi_\nu$ is a solution of the time independent Schr\"{o}dinger equation 
with 
\begin{equation}\label{alpha}
	E=\frac{\hbar^2}{2m}\frac{\nu^2}{4}\quad \text{and}\quad 
	\alpha=\frac{\hbar^2}{2m}f(\nu).
\end{equation}

\noindent {\bf Behavior of the function $f$.}
The last expression for $f$ implies that $f'(\nu)>0$ for $\nu>0$ (and $f'(0)=0$),
because 
$$
\left(-x\cot x\right)'=\frac{x-\,\frac{1}{2}\sin(2x)}{\sin^2x}>0\ \text{for}\ x>0.
$$
We note that
$$
f(\nu)\ \text{is not defined}\ \Leftrightarrow\ \nu=\underline{\nu}_k
\vee
\nu=\overline{\nu}_l,
\ \text{with}\ k,l\in\mathbb{N}.
$$
The expression~\eqref{alternative} for $f$, implies that
as $\nu$ approaches a point where $f$ is not defined from the 
left, then $f(\nu)\nearrow+\infty$; as $\nu$ approaches a point where $f$ is not defined from the 
right, then $f(\nu)\searrow-\infty$:
\begin{eqnarray*}
	\lim_{\nu\nearrow\underline{\nu}_k}f(\nu)=+\infty&\text{and}
	&\lim_{\nu\nearrow\overline{\nu}_l}f(\nu)=+\infty,\\
	\lim_{\nu\searrow\underline{\nu}_k}f(\nu)=-\infty&\text{and}
	&\lim_{\nu\searrow\overline{\nu}_l}f(\nu)=-\infty,
\end{eqnarray*}
whether or not $\underline{\nu}_k=\overline{\nu}_l$.

Let us determine the zeros of $f$:
$$
f(\nu)=0\ \Rightarrow\ \frac{\nu}{2}L=n\pi\ \Leftrightarrow\
\nu_n:=\frac{2n\pi}{L},\ \text{for some}\ n\in
\mathbb{N}.
$$
However, for the reciprocal implication to hold we must make sure
that $\nu$ belongs to the domain of $f$:
\begin{itemize}
	\item
If there exists $(k,l)\in\mathbb{N}^2$ with $k\geq l$ (as $x_0\geq 0$) such that
$n=k+l$ and such that~\eqref{pq} is satisfied (for the given $x_0$),
then $\nu_n=\underline{\nu}_k=\overline{\nu}_l$ and $f$ is not defined at $\nu_n$.
\item
If $\nu_n=\underline{\nu}_k$ for some $k\in\left[\frac{n}{2},n\right)$, i.e.
$$
\frac{2n\pi}{L}=\frac{2k\pi}{\frac{L}{2}+x_0}\
\Leftrightarrow\ x_0=\frac{L}{2}\left(\frac{2k}{n}-1\right),
$$
and we let $l=n-k$, then
$$
\overline{\nu}_l=\frac{2l\pi}{\frac{L}{2}-x_0}=\frac{2n\pi}{L}=\nu_n=\underline{\nu}_k.
$$
So, we fall into the previous situation and $f$ is not defined at $\nu_n$.
\item
Similarly, if $\nu_n=\overline{\nu}_l$ for some $l\in\left(1,\frac{n}{2}\right]$, 
and we let $k=n-l$, then
$\underline{\nu}_k=\overline{\nu}_l$. Again, $f$ is not defined at $\nu_n$.
\end{itemize}
In conclusion, suppose that $x_0=\frac{p}{q}\frac{L}{2}$. Take 
$(k_0,l_0)\in\mathbb{N}^2$ to be the real positive multiple of $\left(\frac{p+q}{2},\frac{q-p}{2}\right)$
closest to the origin.
Then $f$ is not defined for $\nu$ a natural multiple of $\frac{2(k_0+l_0)\pi}{L}$.

\begin{figure}[ht!]
	\begin{psfrags}
		\psfrag{n}{$\nu$}
		\psfrag{f}{\!\!\!\!\!$f(\nu)$}
		\centering
		\includegraphics[scale=1]{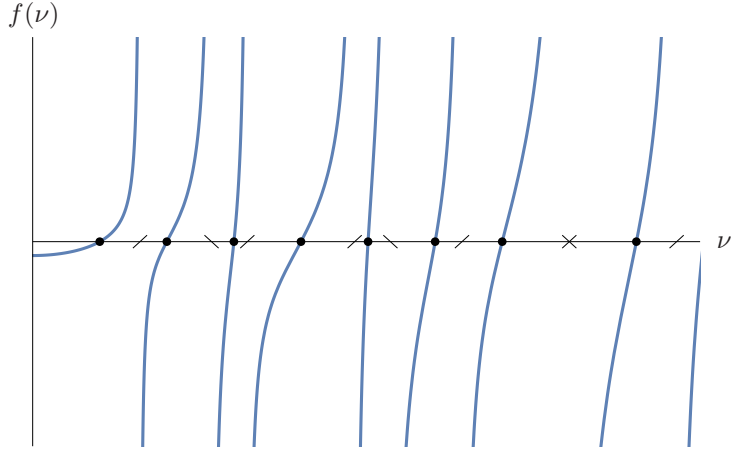}
		\caption{Sketch of the graph of $f$ for $x_0 = \frac{L}{8}$.
			The values $\underline{\nu}_k$ are indicated with a~$\slash$,
			the values $\overline{\nu}_l$ are indicated with a~$\backslash$\,,
			and the values of $\nu_n$ 
			which are zeros of~$f$
			are indicated with a~$\mbox{\tiny $\bullet$}$\,.}
		\label{fig1}
	\end{psfrags}
\end{figure}

\noindent {\bf Example.} Consider $x_0=\frac{1}{4}\frac{L}{2}$. We may take
$p=1$ and $q=4$. The real positive multiple of $\left(\frac{5}{2},\frac{3}{2}\right)$
in $\mathbb{N}^2$ closest to the origin is $(5,3)$. 
The function $f$ 
is zero for $\nu$ of the form $\frac{2n\pi}{L}$,
except if $n$ is divisible by $8$. In Figure~\ref{fig1},
we sketch the graph of $f$ (for this choice of $x_0$).
One might consider superimposing on this plot the vertical line 
crossing the $\nu$ axis at $\nu_8$, since $\Phi_{\nu_8}$ is a 
	solution of the Schr\"{o}dinger equation for any value of $\alpha$.

\subsection{The limit of $\frac{\Psi_\nu}{\rho_\nu}$ as $\nu$ approaches
$\underline{\nu}_k$ or $\overline{\nu}_l$}
\par\ 
\newline
\subsubsection{Case $\nu\to\underline{\nu}_k=\overline{\nu}_l$.}
We wish to calculate the limit of
the solution $\frac{\Psi_\nu}{\rho_\nu}$ as $\nu$ approaches $\underline{\nu}_k=\overline{\nu}_l$. Let $\hat{\nu}$ be equal to the common
value, $\hat{\nu}=\underline{\nu}_k=\overline{\nu}_l$.
We denote by $\Upsilon_{\hat{\nu}}$ the limit function,
\begin{equation}\label{upsilon}
\Upsilon_{\hat{\nu}}(x):=
\lim_{\nu\to\hat{\nu}=\underline{\nu}_k=\overline{\nu}_l}\frac{\Psi_\nu(x)}{\rho_\nu}.
\end{equation}
Before we actually calculate this limit, we make an auxiliary computation
which allows us to simplify its expression. As
\begin{eqnarray}
	\lim_{\nu\to\underline{\nu}_k=\overline{\nu}_l}\frac{\sin\left(\frac{\nu}{2}
		\left(\frac{L}{2}-x_0\right)\right)}{\sin\left(\frac{\nu}{2}
		\left(\frac{L}{2}+x_0\right)\right)}&=&\frac{\frac{L}{2}-x_0}{\frac{L}{2}+x_0}
	\cdot
	\frac{\cos\left(\frac{\nu}{2}
		\left(\frac{L}{2}-x_0\right)\right)}{\cos\left(\frac{\nu}{2}
		\left(\frac{L}{2}+x_0\right)\right)}
		\label{the_point}
		\\
	&=&\frac{l}{k}
	\frac{\cos\left(l\pi\right)}{\cos\left(k\pi\right)}=(-1)^{l-k}\frac{l}{k}
	\nonumber
\end{eqnarray}
(recall~\eqref{lk}), using~\eqref{norm},  we have
\begin{eqnarray}
	\lim_{\nu\to\underline{\nu}_k=\overline{\nu}_l}
	\frac{\rho_\nu}{\left|\sin
		\left(\frac{\nu}{2}\left(\frac{L}{2}+x_0\right)\right)\right|}
	&=&\sqrt{\frac{1}{2}\left(
		\left(\frac{L}{2}+x_0\right)\frac{l^2}{k^2}+\left(\frac{L}{2}-x_0\right)
		\right)}
		\label{point_the}
		\\
	&=&\sqrt{\frac{L}{2}}\sqrt{\frac{l}{k}}.\nonumber
\end{eqnarray}
So, a simpler expression for $\Upsilon_{\hat{\nu}}$ is
$$
\Upsilon_{\hat{\nu}}(x)=\sqrt{\frac{2}{L}}\sqrt{\frac{k}{l}}
\lim_{\nu\to\hat{\nu}=\underline{\nu}_k=\overline{\nu}_l}\frac{\Psi_\nu(x)}{
	\left|\sin\left(\frac{\nu}{2}\left(\frac{L}{2}+x_0\right)\right)\right|}.
$$
Recalling that the sign of $\sin\left(\frac{\nu}{2}\left(\frac{L}{2}+x_0\right)\right)$
is equal to $(-1)^{\left\lfloor\frac{\nu}{2}\left(\frac{L}{2}+x_0\right)\frac{1}{\pi}
	\right\rfloor}$
and recalling~\eqref{kmaisl},
we obtain
\begin{eqnarray*}
	\Upsilon_{\hat{\nu}}(x)&=&
	\begin{cases}
		(-1)^{l-k}\sqrt{\frac{2}{L}}\sqrt{\frac{l}{k}} \sin\left(k\pi\frac{\frac{L}{2} + x}{\frac{L}{2} + x_0}\right), &
		\text{for}\ -\frac{L}{2}<x<x_0,\\
		\sqrt{\frac{2}{L}}\sqrt{\frac{k}{l}}\sin\left(l\pi\frac{\frac{L}{2} - x}{\frac{L}{2} - x_0}\right), & \text{for}\ x_0<x<\frac{L}{2},
	\end{cases}\\
	&=&\begin{cases}
		(-1)^{l-k}\sqrt{\frac{2}{L}}\sqrt{\frac{l}{k}} \sin\left(\frac{(k+l)\pi}{L}\left(\frac{L}{2} + x\right)\right), &
		\text{for}\ -\frac{L}{2}<x<x_0,\\
		\sqrt{\frac{2}{L}}\sqrt{\frac{k}{l}}\sin\left(\frac{(k+l)\pi}{L}\left(\frac{L}{2} - x\right)\right), & \text{for}\ x_0<x<\frac{L}{2},
	\end{cases}\\
	&=&\begin{cases}
	-\sqrt{\frac{2}{L}}\sqrt{\mathpzc{q}}\, \sin\left(\frac{\hat{\nu}}{2}\left(\frac{L}{2} - x\right)\right), &
	\text{for}\ -\frac{L}{2}<x<x_0,\\
	\sqrt{\frac{2}{L}}\frac{1}{\sqrt{\mathpzc{q}}}\,\sin\left(\frac{\hat{\nu}}{2}\left(\frac{L}{2} - x\right)\right), & \text{for}\ x_0<x<\frac{L}{2},
	\end{cases}
	\\
	&=&\begin{cases}
		-\sqrt{\mathpzc{q}}\,
		\Phi_{\hat{\nu}}(x), &
		\text{for}\ -\frac{L}{2}<x<x_0,\\
		\frac{1}{\sqrt{\mathpzc{q}}}\,
		\Phi_{\hat{\nu}}(x), & \text{for}\ x_0<x<\frac{L}{2},
	\end{cases}
\end{eqnarray*}	
The amplitude of the oscillations on the left of $x_0$ is $\mathpzc{q}$
times the amplitude of the oscillations on the right of $x_0$.

\begin{rmk}\label{nice}
	If $\nu=\underline{\nu}_k=\overline{\nu}_l$, then we have shown
	$\nu=\nu_{k+l}$.
For a particle in the state $\Phi_{\nu_{k+l}}$, the ratio of the probability
of finding the particle on the right of $x_0$ over the probability of finding the particle on the left of $x_0$
is equal to ratio of the length of the interval on the right of $x_0$
over the length of the interval on the left of $x_0$:
	$$
	\frac{
		\int_{x_0}^{\frac{L}{2}}\Phi_{\nu_{k+l}}^2(x)\,dx}
	{\int_{-\frac{L}{2}}^{x_0}\Phi_{\nu_{k+l}}^2(x)\,dx}
	=\frac{\frac{1}{L}\left(\frac{L}{2}-x_0\right)}
	{\frac{1}{L}\left(\frac{L}{2}+x_0\right)}=
	\mathpzc{q}.
	$$
For a particle in the limit state $\Upsilon_{\hat{\nu}}$
the ratio of the probability of finding the particle
on the right of $x_0$ over the probability of finding the particle on the left of $x_0$ is the inverse
of the one of for the particle in the state~$\Phi_{\nu_{k+l}}$:
	\begin{equation}\label{prob}
\frac{
	\int_{x_0}^{\frac{L}{2}}\Upsilon_{\hat{\nu}}^2(x)\,dx}
{\int_{-\frac{L}{2}}^{x_0}\Upsilon_{\hat{\nu}}^2(x)\,dx}=
\frac{
	\frac{1}{\mathpzc{q}}\int_{x_0}^{\frac{L}{2}}\Phi_{\nu_{k+l}}^2(x)\,dx}
{\mathpzc{q}\int_{-\frac{L}{2}}^{x_0}\Phi_{\nu_{k+l}}^2(x)\,dx}
=\frac{1}{\mathpzc{q}}.
\end{equation}
\end{rmk}

\begin{rmk}\label{uniform}
	Given $\tilde{\nu}\in\mathbb{R}$,
	suppose that $\nu\to\tilde{\nu}$. It is certainly true that
	$\sin\left(\frac{\nu}{2}\left(\frac{L}{2}+\,\cdot\,\right)\right)$
	converges in $C^1$ to
	$\sin\left(\frac{\tilde{\nu}}{2}\left(\frac{L}{2}+\,\cdot\,\right)\right)$,
	and that
	$\sin\left(\frac{\nu}{2}\left(\frac{L}{2}-\,\cdot\,\right)\right)$
	converges in $C^1$ to
	$\sin\left(\frac{\tilde{\nu}}{2}\left(\frac{L}{2}-\,\cdot\,\right)\right)$.
	The limits in~\eqref{the_point} and\/~\eqref{point_the}, are limits of
	sequences of real numbers, not limits of functions. For these two reasons,
	the convergence of $\frac{1}{\rho_\nu}\Psi_\nu$ to
	$\Upsilon_{\hat{\nu}}$ is uniform in $\left[-\,\frac{L}{2},\frac{L}{2}\right]$,
	is $C^1$ in $\left[-\,\frac{L}{2},x_0\right]$, and is $C^1$ in
	$\left[x_0,\frac{L}{2}\right]$.
\end{rmk}

In order to obtain the limit equation satisfied by 
$\Upsilon_{\hat{\nu}}$ we calculate the derivatives
\begin{eqnarray*}
\Upsilon'_{\hat{\nu}}(x_0^-)&=&
\sqrt{\frac{2}{L}}\sqrt{\mathpzc{q}}\frac{\hat{\nu}}{2}\,(-1)^{l-k}(-1)^k\ =\ \frac{1}{\sqrt{2L}}\sqrt{\mathpzc{q}}\hat{\nu}(-1)^l,\\
\Upsilon'_{\hat{\nu}}(x_0^+)&=&
-\sqrt{\frac{2}{L}}\frac{1}{\sqrt{\mathpzc{q}}}\frac{\hat{\nu}}{2}(-1)^l\ =\
-\frac{1}{\sqrt{2L}}\frac{1}{\sqrt{\mathpzc{q}}}\hat{\nu}(-1)^l.
\end{eqnarray*}
In particular, we have that
$$
\Upsilon_{\hat{\nu}}(x_0^-)=
-\,\mathpzc{q}\,\Upsilon_{\hat{\nu}}(x_0^+).
$$
Their difference is
$$
\Upsilon'_{\hat{\nu}}(x_0^+)-\Upsilon'_{\hat{\nu}}(x_0^-)
=\frac{1}{\sqrt{2L}}\frac{\mathpzc{q}+1}{\sqrt{\mathpzc{q}}}\hat{\nu}(-1)^{l-1}.
$$
It depends, of course, on 
our normalization $\Bigl(\|\Upsilon_{\hat{\nu}}\|_{L^2\left(
	-\,\frac{L}{2},\frac{L}{2}\right)}=1\Bigr.$ and 
$\Upsilon_{\hat{\nu}}$ positive in a left 
neighborhood of $\Bigl.\frac{L}{2}\Bigr)$.
Let
\begin{equation}\label{leitao}
\kappa_{\hat{\nu}}:=
\frac{\hbar^2}{2m}\frac{1}{\sqrt{2L}}\frac{\mathpzc{q}+1}{\sqrt{\mathpzc{q}}}\hat{\nu}(-1)^{l-1}=
\frac{\hbar^2}{2m}
\frac{\sqrt{2L}}{\sqrt{L+2x_0}\sqrt{L-2x_0}}
\hat{\nu}(-1)^{l-1}.
\end{equation}
The function $\Upsilon_{\hat{\nu}}$ is a weak solution of the limit equation
\begin{equation}\label{eq1_alt}
	\begin{cases}
		-\frac{\hbar^{2}}{2 m} \frac{d^2\Upsilon_{\hat{\nu}}}{dx^2} +\sigma \kappa_{\hat{\nu}}
		\|\Upsilon_{\hat{\nu}}\|_{L^2\left(-\,\frac{L}{2},\frac{L}{2}\right)}
		\delta(x - x_0)
		 = E_{\hat{\nu}} \Upsilon_{\hat{\nu}}, \\
		\Upsilon_{\hat{\nu}}\left(-\,\frac{L}{2}\right)= \Upsilon_{\hat{\nu}}\left(\frac{L}{2}\right) = 0,
	\end{cases}  
\end{equation}
where $\sigma$ is the sign of $\Upsilon_{\hat{\nu}}$
in a left neighborhood of $\frac{L}{2}$.

Let us calculate the Fourier series of $\Upsilon_{\hat{\nu}}$,
$$
\Upsilon_{\hat{\nu}}(x)=\sum_{m=1}^\infty a_m\Phi_{\nu_m}(x).
$$
This series converges uniformly since $\Upsilon_{\hat{\nu}}$ is piecewise $C^1$.
Testing~\eqref{eq1_alt} with $\Phi_{\nu_m}$, we obtain
	$$ 
	-\frac{\hbar^{2}}{2 m} \int_{-\frac{L}{2}}^{\frac{L}{2}}
	\Upsilon_{\hat{\nu}}(x)\frac{d^2\Phi_{\nu_m}}{d{x}^2}(x)\,dx + \kappa_{\hat{\nu}} \int_{-\frac{L}{2}}^{\frac{L}{2}}\delta(x - x_0) \Phi_{\nu_m}(x)\,dx = E_{\hat{\nu}}\int_{-\frac{L}{2}}^{\frac{L}{2}} \Upsilon_{\hat{\nu}}(x)\Phi_{\nu_m}(x)\,dx.
	$$ 
	Let $p:=k+l$, so that $\hat{\nu}=\nu_p$.
Note that
$-\Phi_{\nu_m}''=\left(\frac{\nu_m}{2}\right)^2\Phi_{\nu_m}=\frac{\pi^2m^2}{L^2}
\Phi_{\nu_m}$, 
$E_{\hat{\nu}}=\frac{\hbar^2}{2m}\left(\frac{\hat{\nu}}{2}\right)^2=
\frac{\hbar^2}{2m}\frac{\pi^2p^2}{L^2}$, and $\kappa_{\hat{\nu}}$ has a factor
$\frac{\hbar^2}{2m}$. Hence, dividing by this factor, we obtain
\begin{equation}\label{am}
\frac{\pi^2m^2}{L^2}a_m+\frac{\sqrt{2L}}{\sqrt{L+2x_0}\sqrt{L-2x_0}}
\frac{2\pi p}{L}(-1)^{l-1}\Phi_{\nu_m}(x_0)=\frac{\pi^2 p^2}{L^2}a_m.
\end{equation}
The $l$ in this expression is the $l$ in~\eqref{leitao},
the number of oscillations of $\Upsilon_{\hat{\nu}}$ on the right of $x_0$.
Solving for $a_m$, we get
\begin{eqnarray*}
a_m&=&(-1)^l\frac{2\sqrt{2}pL^{\frac{3}{2}}}{\pi\sqrt{L^2-4x_0^2}}
\frac{\Phi_{\nu_m}(x_0)}{m^2-p^2}\\
&=&\cos\left(\frac{p\pi}{L}\left(\frac{L}{2}-x_0\right)\right)
\frac{2\sqrt{2}pL^{\frac{3}{2}}}{\pi\sqrt{L^2-4x_0^2}}
\frac{\Phi_{\nu_m}(x_0)}{m^2-p^2},
\end{eqnarray*}
for all $m\in\mathbb{N}\setminus\{p\}$. The value of $a_p$ is
\begin{eqnarray*}
a_p&=&\int_{-\,\frac{L}{2}}^{\frac{L}{2}}\Upsilon_{\hat{\nu}}(x)\Phi_{\nu_p}(x)\,dx\\
&=&-\sqrt{\mathpzc{q}}\int_{-\,\frac{L}{2}}^{x_o}\Phi_{\nu_p}^2(x)\,dx+
\frac{1}{\sqrt{\mathpzc{q}}}\int_{x_0}^{\frac{L}{2}}\Phi_{\nu_p}^2(x)\,dx\\
&=&\frac{2}{L}\left(
-\sqrt{\frac{\frac{L}{2}-x_0}{\frac{L}{2}+x_0}}\frac{1}{2}\left(\frac{L}{2}+x_0\right)
+\sqrt{\frac{\frac{L}{2}+x_0}{\frac{L}{2}-x_0}}\frac{1}{2}\left(\frac{L}{2}-x_0\right)
\right)\\
&=& 0.
\end{eqnarray*}
This shows that
\begin{equation}\label{Fourier}
\Upsilon_{\nu_p}(x)=c_{x_0,p}
\sum_{m=1}^\infty
\frac{\Phi_{\nu_m}(x_0)}{m^2-p^2}\Phi_{\nu_m}(x),
\end{equation}
with the convention that
$$
m=p\ \ \ \Longrightarrow\ \ \ \frac{\Phi_{\nu_m}(x_0)}{m^2-p^2}=0,
$$
and with
$$
c_{x_0,p}:=\cos\left(\frac{p\pi}{L}\left(\frac{L}{2}-x_0\right)\right)
\frac{2\sqrt{2}pL^{\frac{3}{2}}}{\pi\sqrt{L^2-4x_0^2}}.
$$

The expression~\eqref{Fourier} shows that the function $\Upsilon_{\nu_p}$ does not have modes that vanish at $x_0$,
i.e.\ $\Phi_{\nu_m}(x_0)=0$ implies $a_m\Phi_{\nu_m}\equiv0$
(as $a_m=0$). This is a consequence of the symmetry of the Schr\"{o}dinger operator.
Indeed, fix an $n$ such that $\Phi_{\nu_n}(x_0)=0$. Then,
$\Phi_{\nu_n}$ is an eigenfunction of the Schr\"{o}dinger operator
for all values of $\alpha$. For each $\alpha_0$, with $|\alpha_0|$ large,
the function $\frac{\Psi_{\nu}}{\rho_\nu}$, with $\nu$ close to $\nu_p$
such that $\frac{\hbar^2}{2m}f(\nu)=\alpha_0$, is $L^2$ orthogonal to 
$\Phi_{\nu_n}$ (even in the case that $n=p$), as $\nu$ is different from
$\nu_n$ (and hence $\frac{\Psi_{\nu}}{\rho_\nu}$ and $\Phi_{\nu_n}$ correspond to 
different eigenvalues of the Schr\"{o}dinger operator with parameter $\alpha_0$). Letting $|\alpha_0|\to+\infty$, $\nu$ goes to $\nu_p$ and so we conclude that
$\Upsilon_{\nu_p}$
is $L^2$ orthogonal to $\Phi_{\nu_n}$. This argument shows, as do the computations
preceding~\eqref{Fourier}, that, provided $\Phi_{\nu_n}(x_0)=0$,
$\Phi_{\nu_n}$ is $L^2$ orthogonal to $\Upsilon_{\nu_p}$, even in the case that
$n=p$ and the two eigenfunctions ($\Phi_{\nu_p}$ and $\Upsilon_{\nu_p}$) correspond to the same eigenvalue
(i.e.\ have the same energy).

\subsubsection{Case $\nu\to\underline{\nu}_k\neq\overline{\nu}_l$.} 

We have
\begin{eqnarray*}
	\underline{\Upsilon}_{\underline{\nu}_k}^-(x)&:=&\lim_{\nu\nearrow\underline{\nu}_k\neq\overline{\nu}_l}\frac{\Psi_\nu(x)}
	{\rho_\nu}\\
&=&\frac{2}{\sqrt{L+2x_0}}\lim_{\nu\nearrow\underline{\nu}_k\neq\overline{\nu}_l}\frac{\Psi_\nu(x)}
{\left|\sin\left(\frac{\nu}{2}\left(\frac{L}{2}-x_0\right)\right)\right|}\\
&=&
\begin{cases}
	(-1)^{\left\lfloor\frac{\underline{\nu}_k}{2}\left(\frac{L}{2}-x_0\right)
		\frac{1}{\pi}\right\rfloor+k-1}
	\frac{2}{\sqrt{L+2x_0}}
	 \sin\left(k\pi\frac{\frac{L}{2} + x}{\frac{L}{2} + x_0}\right), &
	\text{for}\ -\frac{L}{2}<x<x_0,\\
	0, & \text{for}\ x_0<x<\frac{L}{2},
\end{cases}\\
&=&
\begin{cases}
	(-1)^{\left\lfloor \frac{kL}{\frac{L}{2}+x_0}
		\right\rfloor-1}
	\frac{2}{\sqrt{L+2x_0}}
	\sin\left(k\pi\frac{\frac{L}{2} + x}{\frac{L}{2} + x_0}\right), &
	\text{for}\ -\frac{L}{2}<x<x_0,\\
	0, & \text{for}\ x_0<x<\frac{L}{2}.
\end{cases}
\end{eqnarray*}
\begin{eqnarray*}
	\underline{\Upsilon}_{\underline{\nu}_k}^+(x)&:=&\lim_{\nu\searrow\underline{\nu}_k\neq\overline{\nu}_l}\frac{\Psi_\nu(x)}
	{\rho_\nu}
	\ =\ -\underline{\Upsilon}_{\underline{\nu}_k}^-(x).
\end{eqnarray*}
\begin{rmk}\label{sign}
The difference of the signs of $\underline{\Upsilon}_{\underline{\nu}_k}^-$ and
$\underline{\Upsilon}_{\underline{\nu}_k}^+$ is due to our normalization,
we require our solutions to be positive in a left neighborhood of $\frac{L}{2}$.
Remember that $\Psi_\nu$ is only defined for $\nu\not\in\mathcal{P}$.
If we redefined our~$\Psi_\nu$ to $-\Psi_\nu$ each time
we crossed a point in $\underline{\mathcal{P}}\setminus\mathcal{K}$, then
we would obtain 
$\underline{\Upsilon}_{\underline{\nu}_k}^-=\underline{\Upsilon}_{\underline{\nu}_k}^+$.
\end{rmk}
\begin{rmk}
	The convergence of $\frac{1}{\rho_\nu}\Psi_\nu$ to
	$\underline{\Upsilon}_{\underline{\nu}_k}^-$, as $\nu\nearrow\underline{\nu}_k\neq\overline{\nu}_l$,
	is uniform in $\left[-\,\frac{L}{2},\frac{L}{2}\right]$,
	is $C^1$ in $\left[-\,\frac{L}{2},x_0\right]$, and is $C^1$ in
	$\left[x_0,\frac{L}{2}\right]$. The convergence of $\frac{1}{\rho_\nu}\Psi_\nu$ to
	$\underline{\Upsilon}_{\underline{\nu}_k}^+$, as $\nu\searrow\underline{\nu}_k\neq\overline{\nu}_l$,
	is uniform in $\left[-\,\frac{L}{2},\frac{L}{2}\right]$,
	is $C^1$ in $\left[-\,\frac{L}{2},x_0\right]$, and is $C^1$ in
	$\left[x_0,\frac{L}{2}\right]$. The justification is the same as the one
	of\/ {\rm Remark~\ref{uniform}}.
\end{rmk}

Letting
$$
\underline{\kappa}_{\underline{\nu}_k}:=
\underline{\kappa}_{\underline{\nu}_k}^-:=-\,
\frac{\hbar^2}{2m}\underline{\Upsilon}_{\underline{\nu}_k}^{-\prime}(x_0^-)
=
\frac{\hbar^2}{2m}\frac{1}{\sqrt{L+2x_0}}\underline{\nu}_k
(-1)^{\left\lfloor\frac{\underline{\nu}_k}{2}\left(\frac{L}{2}-x_0\right)
	\frac{1}{\pi}\right\rfloor},
$$
$$
\underline{\kappa}_{\underline{\nu}_k}^+:=-
\underline{\kappa}_{\underline{\nu}_k}^-,
$$
the limit equation is
\begin{equation}\label{eq1_alt2}
	\begin{cases}
		-\frac{\hbar^{2}}{2 m} \frac{d^2\underline{\Upsilon}_{\underline{\nu}_k}}{dx^2} +\sigma \underline{\kappa}_{\underline{\nu}_k}
		\|\underline{\Upsilon}_{\underline{\nu}_k}\|_{L^2\left(-\,\frac{L}{2},\frac{L}{2}\right)}
		\delta(x - x_0)
		= E_{\underline{\nu}_k} \underline{\Upsilon}_{\underline{\nu}_k}, \\
		\underline{\Upsilon}_{\underline{\nu}_k}\left(-\,\frac{L}{2}\right)= \underline{\Upsilon}_{\underline{\nu}_k}\left(\frac{L}{2}\right) = 0,
	\end{cases}  
\end{equation}
where $\sigma$ is the sign of $\underline{\Upsilon}_{\underline{\nu}_k}$
multiplied by $(-1)^{\left\lfloor \frac{kL}{\frac{L}{2}+x_0}
	\right\rfloor-1}
$
in a right neighborhood of~$-\,\frac{L}{2}$.

When we calculate the Fourier series for 
$\underline{\Upsilon}_{\underline{\nu}_k}:=\underline{\Upsilon}_{\underline{\nu}_k}^-$,
equality~\eqref{am} is replaced by
\begin{equation}\label{am2}
	\frac{\pi^2m^2}{L^2}a_m+\frac{1}{\sqrt{L+2x_0}}
	\frac{2\pi k}{\frac{L}{2}+x_0}
	(-1)^{\left\lfloor\frac{\underline{\nu}_k}{2}\left(\frac{L}{2}-x_0\right)
		\frac{1}{\pi}\right\rfloor}\Phi_{\nu_m}(x_0)=\frac{\pi^2 k^2}{\left(\frac{L}{2}+x_0\right)^2}a_m.
\end{equation}
Solving for $a_m$, we get
$$
a_m=(-1)^{1+\left\lfloor k\left(\frac{\frac{L}{2}-x_0}{\frac{L}{2}+x_0}\right)
	\right\rfloor}
	\frac{kL^2\sqrt{L+2x_0}}{\pi}
	\frac{\Phi_{\nu_m}(x_0)}{\left(\frac{L}{2}+x_0\right)^2m^2-L^2k^2}.
$$
This shows that
\begin{equation}\label{Fourier_2}
	\underline{\Upsilon}_{\underline{\nu}_k}(x)=\underline{c}_{x_0,k}
	\sum_{m=1}^\infty
	\frac{\Phi_{\nu_m}(x_0)}{\left(\frac{L}{2}+x_0\right)^2m^2-L^2k^2}\Phi_{\nu_m}(x),
\end{equation}
with
$$
\underline{c}_{x_0,k}:=(-1)^{1+\left\lfloor k\left(\frac{\frac{L}{2}-x_0}{\frac{L}{2}+x_0}\right)
	\right\rfloor}
\frac{kL^2\sqrt{L+2x_0}}{\pi}.
$$

\subsubsection{Case $\nu\to\overline{\nu}_l\neq\underline{\nu}_k$.}

We have
\begin{eqnarray*}	\overline{\Upsilon}_{\overline{\nu}_l}(x)&:=&\lim_{\nu\to\overline{\nu}_l\neq\underline{\nu}_k}\frac{\Psi_\nu(x)}
	{\rho_\nu}\\
&=&\frac{2}{\sqrt{L-2x_0}}
\lim_{\nu\to\overline{\nu}_l\neq\underline{\nu}_k}\frac{\Psi_\nu(x)}
{\left|\sin\left(\frac{\nu}{2}\left(\frac{L}{2}+x_0\right)\right)\right|}\\
&=&
\begin{cases}
	0, &
	\text{for}\ -\frac{L}{2}<x<x_0,\\
	 \frac{2}{\sqrt{L-2x_0}}\sin\left(l\pi\frac{\frac{L}{2} - x}{\frac{L}{2} - x_0}\right), & \text{for}\ x_0<x<\frac{L}{2}.
\end{cases}
\end{eqnarray*}

\begin{rmk}
	The convergence of $\frac{1}{\rho_\nu}\Psi_\nu$ to
	$\overline{\Upsilon}_{\overline{\nu}_l}$, as $\nu\to\overline{\nu}_l\neq\underline{\nu}_k$,
	is uniform in $\left[-\,\frac{L}{2},\frac{L}{2}\right]$,
	is $C^1$ in $\left[-\,\frac{L}{2},x_0\right]$, and is $C^1$ in
	$\left[x_0,\frac{L}{2}\right]$. 
	Again, the justification is the same as the one
	of\/ {\rm Remark~\ref{uniform}}.
\end{rmk}

Letting
$$
\overline{\kappa}_{\overline{\nu}_l}:=\frac{\hbar^2}{2m}
\overline{\Upsilon}'_{\overline{\nu}_l}(x_0^+)
=\frac{\hbar^2}{2m}\frac{1}{\sqrt{L-2x_0}}\overline{\nu}_l(-1)^{l-1},
$$
the limit equation is
\begin{equation}\label{eq1_alt3}
	\begin{cases}
		-\frac{\hbar^{2}}{2 m} \frac{d^2\overline{\Upsilon}_{\overline{\nu}_l}}{dx^2} +\sigma \overline{\kappa}_{\overline{\nu}_l}
		\|\overline{\Upsilon}_{\overline{\nu}_l}\|_{L^2\left(-\,\frac{L}{2},\frac{L}{2}\right)}
		\delta(x - x_0)
		= E_{\overline{\nu}_l} \overline{\Upsilon}_{\overline{\nu}_l}, \\
		\overline{\Upsilon}_{\overline{\nu}_l}\left(-\,\frac{L}{2}\right)= \overline{\Upsilon}_{\overline{\nu}_l}\left(\frac{L}{2}\right) = 0,
	\end{cases}  
\end{equation}
where $\sigma$ is the sign of $\overline{\Upsilon}_{\overline{\nu}_l}$
in a left neighborhood of $\frac{L}{2}$.

When we calculate the Fourier series for $\overline{\Upsilon}_{\overline{\nu}_l}$,
equality~\eqref{am2} is replaced by
$$
	\frac{\pi^2m^2}{L^2}a_m+\frac{1}{\sqrt{L-2x_0}}
	\frac{2\pi l}{\frac{L}{2}-x_0}
	(-1)^{l-1}\Phi_{\nu_m}(x_0)=\frac{\pi^2 l^2}{\left(\frac{L}{2}-x_0\right)^2}a_m.
$$
Solving for $a_m$, we get
$$
a_m=(-1)^{l}
\frac{lL^2\sqrt{L-2x_0}}{\pi}
\frac{\Phi_{\nu_m}(x_0)}{\left(\frac{L}{2}-x_0\right)^2m^2-L^2l^2}.
$$
This shows that
\begin{equation}\label{Fourier_3}
\overline{\Upsilon}_{\overline{\nu}_k}(x)=\overline{c}_{x_0,l}
\sum_{m=1}^\infty
\frac{\Phi_{\nu_m}(x_0)}{\left(\frac{L}{2}-x_0\right)^2m^2-L^2l^2}\Phi_{\nu_m}(x),
\end{equation}
with
$$
\overline{c}_{x_0,l}:=(-1)^{l}
\frac{lL^2\sqrt{L-2x_0}}{\pi}.
$$

\subsection{Partition of the positive real line by $\mathcal{P}$} 

Note that
\begin{equation}\label{G}
\nu_1\in(0,\underline{\nu}_1).
\end{equation}
 We start with $\nu=\nu_n=\frac{2n\pi}{L}$, with $n$ a natural number greater than or equal to $2$, and $\alpha=0$, so that
$\Psi_{\nu_n}=\Phi_{\nu_n}$ is a solution of~\eqref{eq1}.
Observe that $\underline{\nu}_1\leq\nu_2$.
The set $\mathcal{P}$
partitions $[\underline{\nu}_1,+\infty)$ into a family of open subintervals.
We wish to identify the open subinterval containing $\nu_n$, or the
point of $\mathcal{P}$ equal to $\nu_n$.
Let
\begin{eqnarray*}
	\hat{k}&=&\left\lfloor\frac{\nu_n}{\underline{\nu}_1}\right\rfloor\ =\ 
	\left\lfloor n\left(\frac{1}{2}+\frac{x_0}{L}\right)\right\rfloor,\\
	\hat{l}&=&\left\lfloor\frac{\nu_n}{\overline{\nu_1}}\right\rfloor\ =\ 
	\left\lfloor n\left(\frac{1}{2}-\frac{x_0}{L}\right)\right\rfloor.
\end{eqnarray*}
Obviously, $\hat{k}\geq\hat{l}$. The value $\max\{\hat{k}\underline{\nu}_1,\hat{l}\overline{\nu}_1\}$ is the
greatest element of the set $\mathcal{P}$
which is less than or equal to $\nu_n$. Several cases might occur:
\begin{enumerate}[(A)]
	\item $\max\{\hat{k}\underline{\nu}_1,\hat{l}\overline{\nu}_1\}=\nu_n$.
We note that
	$$
	\underline{\nu}_{\hat{k}}=\nu_n\ \Leftrightarrow\ 
	\frac{\frac{L}{2}+x_0}{\hat{k}}=\frac{L}{n}\ \Leftrightarrow\ 
	\frac{\frac{L}{2}-x_0}{n-\hat{k}}=\frac{L}{n}\ \Leftrightarrow\ 
	\overline{\nu}_{n-\hat{k}}=\nu_n.
	$$
	So, if the maximum is equal to $\nu_n$, then in fact
	\begin{equation}\label{Z}
	\nu_n=\underline{\nu}_{k}=\overline{\nu}_{l},
\end{equation}
	where $k=\hat{k}$ and $l=\hat{l}$.
	Hence the solution $\Phi_{\nu_n}$ will be unaffected by the value of $\alpha$.
\item $\max\{\hat{k}\underline{\nu}_1,\hat{l}\overline{\nu}_1\}<\nu_n$.
\begin{enumerate}[(a)]
	\item $\underline{\nu}_{\hat{k}}\neq \overline{\nu}_{\hat{l}}$
 and $\underline{\nu}_{\hat{k}+1}\neq \overline{\nu}_{\hat{l}+1}$.
 It is not possible to have $\nu_n\in
 (\overline{\nu}_{l},\overline{\nu}_{l+1})$,
 with this a subinterval defined by $\mathcal{P}$, because this would force 
 $\underline{\nu}_1=\overline{\nu}_1$, which would imply
 $\underline{\nu}_{\hat{k}}=\overline{\nu}_{\hat{l}}$
 and $\underline{\nu}_{\hat{k}+1}=\overline{\nu}_{\hat{l}+1}$,
 which is ruled out in (a). So,
  there are three possibilities:
\begin{enumerate}[(i)]
	\item 
	\begin{equation}\label{A}
	\nu_n\in(\underline{\nu}_k,\underline{\nu}_{k+1}),
	\end{equation}
	 where $k=\hat{k}$.
	\item 
	\begin{equation}\label{B}
	\nu_n\in(\underline{\nu}_k,\overline{\nu}_{l}),
	\end{equation}
	 where $k=\hat{k}$
	and $l=\hat{l}+1$.
	\item 
	\begin{equation}\label{C}
	\nu_n\in(\overline{\nu}_l,\underline{\nu}_{k}),
	\end{equation}
	 where $l=\hat{l}$
	and $k=\hat{k}+1$.
\end{enumerate}
\item $x_0=0$.   
Then
$\hat{k}=\hat{l}=\left\lfloor\frac{n}{2}\right\rfloor$.
If $n$ is even, we have $\hat{k}=\hat{l}=\frac{n}{2}$,
so $\underline{\nu}_{\frac{n}{2}}=\overline{\nu}_{\frac{n}{2}}$,
whence this common value must be $\nu_n$. This contradicts 
$\max\{\hat{k}\underline{\nu}_1,\hat{l}\overline{\nu}_1\}<\nu_n$.
So, 
\begin{equation}\label{odd} 
	\max\{\hat{k}\underline{\nu}_1,\hat{l}\overline{\nu}_1\}<\nu_n
	\ \,\text{and}\ \, x_0=0
	\ \ \
	\Longrightarrow\ \ \
	n\ \text{has to be odd}
\end{equation}
 (the solution is even) and
\begin{equation}\label{D}
\nu_n\in\left(\underline{\nu}_{\frac{n-1}{2}},\underline{\nu}_{\frac{n+1}{2}}
\right)=
\left(\overline{\nu}_{\frac{n-1}{2}},\overline{\nu}_{\frac{n+1}{2}}
\right)=(\nu_{n-1},\nu_{n+1}).
\end{equation}
This coresponds to the situation studied in~\cite{jviana11}
(see Figure~\ref{line-zero}).

\begin{figure}[ht!]
	\centering
	\begin{psfrags}	
		\psfrag{a}{$\nu$}
		\psfrag{b}{$\underline{\nu}$}
		\psfrag{c}{$\overline{\nu}$}
		\psfrag{d}{\!\!$\underline{\nu}_{\mbox{\tiny 5}}\!\!=\!\overline{\nu}_{\mbox{\tiny 3}}$}
		\includegraphics[scale = 1]{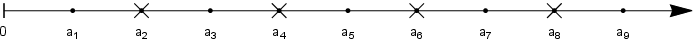}
		\caption{The set $\mathcal{P}$ and the values of $\nu_n$ for 
			$x_0=0$.}
		\label{line-zero}
	\end{psfrags}
\end{figure}

    \item $x_0\neq 0$.
    \begin{enumerate}[(i)]
	\item  $\underline{\nu}_{\hat{k}}=\overline{\nu}_{\hat{l}}$.
   We have $\underline{\nu}_{k+1}<\overline{\nu}_{l+1}$ and
   $\nu_n\in(\underline{\nu}_k,\underline{\nu}_{k+1})=
   (\overline{\nu}_l,\underline{\nu}_{k+1})$, where $k=\hat{k}$ and
   $l=\hat{l}$.
   Since  $\underline{\nu}_{\hat{k}}=\overline{\nu}_{\hat{l}}$, it follows
   that $\nu_{\hat{k}+\hat{l}}$ is this common value.
   Using 
   $\lfloor x+y\rfloor-1\leq\lfloor x\rfloor+\lfloor y\rfloor\leq \lfloor x+y\rfloor$,
   we obtain $n-1\leq\hat{k}+\hat{l}\leq n$, and so
   $\nu_{n-1}\leq \nu_{\hat{k}+\hat{l}}\leq\nu_n$. Then, either
   $\nu_{\hat{k}+\hat{l}}=\nu_{n-1}$ or
   $\nu_{\hat{k}+\hat{l}}=\nu_n$. The second possibility leads to the
   contradiction $\nu_n>\underline{\nu}_{\hat{k}}=\nu_{\hat{k}+\hat{l}}=\nu_n$.
   We conclude that $\nu_{n-1}=\nu_{\hat{k}+\hat{l}}=
   \underline{\nu}_{k}=\overline{\nu}_{l}$:
   \begin{equation}\label{E}
   \nu_n\in(\underline{\nu}_k,\underline{\nu}_{k+1})=
   (\overline{\nu}_l,\underline{\nu}_{k+1})=(\nu_{n-1},\underline{\nu}_{k+1}),
   \end{equation}
    where $k=\hat{k}$ and
   $l=\hat{l}$.
	\item $\underline{\nu}_{\hat{k}+1}=\overline{\nu}_{\hat{l}+1}$. 
	We have
	$\overline{\nu}_{l-1}<\underline{\nu}_{k-1}$
	and
	$ 
	\nu_n\in(\underline{\nu}_{k-1},\underline{\nu}_{k})=
	(\underline{\nu}_{k-1},\overline{\nu}_{l})
	$, 
	 where $k=\hat{k}+1$ and
	$l=\hat{l}+1$. 
	Since  $\underline{\nu}_{\hat{k}+1}=\overline{\nu}_{\hat{l}+1}$, it follows
	that $\nu_{(\hat{k}+1)+(\hat{l}+1)}
	=\nu_{\hat{k}+\hat{l}+2}=\underline{\nu}_{\hat{k}+1}=\overline{\nu}_{\hat{l}+1}$.
	Knowing that
	$\nu_{n+1}\leq \nu_{\hat{k}+\hat{l}+2}\leq\nu_{n+2}$
	(because we saw above that $n-1\leq\hat{k}+\hat{l}\leq n$),
	either
	$\nu_{\hat{k}+\hat{l}+2}=\nu_{n+1}$ or
	$\nu_{\hat{k}+\hat{l}+2}=\nu_{n+2}$. 
	The equality $\nu_{n+2}=\nu_{\hat{k}+\hat{l}+2}
	=\underline{\nu}_{\hat{k}+1}=\underline{\nu}_k$ cannot hold: 
	$\underline{\nu}_k=\underline{\nu}_{k-1}+\frac{2\pi}{\frac{L}{2}+x_0}
	\leq\underline{\nu}_{k-1}+\frac{4\pi}{L}=\underline{\nu}_{\hat{k}}+\frac{4\pi}{L}
	<
	\nu_n+\frac{4\pi}{L}=\nu_{n+2}$.
	We conclude that $\nu_{n+1}=\nu_{\hat{k}+\hat{l}+2}=
	\underline{\nu}_{k}=\overline{\nu}_{l}$:
	\begin{equation}\label{F}
		\nu_n\in(\underline{\nu}_{k-1},\underline{\nu}_{k})=
	(\underline{\nu}_{k-1},\overline{\nu}_{l})=(\underline{\nu}_{k-1},\nu_{n+1}),
	\end{equation}
	 where $k=\hat{k}+1$ and
	$l=\hat{l}+1$.
\end{enumerate}
\end{enumerate}
\end{enumerate}
If $x_0$ is an irrational multiple of $L$, then, obviously,
only~\eqref{A}, \eqref{B} or~\eqref{C} can arise.

\begin{rmk}\label{m}
	The inclusion $\overline{\mathcal{P}}\subset\underline{\mathcal{P}}$ happens
	if and only if $\overline{\nu}_1$ is a natural multiple of
	(say $m$ times) 
	$\underline{\nu}_1$, i.e.\ it happens when $x_0=\frac{m-1}{m+1}\frac{L}{2}$. 
	As
	$$
	\frac{\overline{\nu}_1}{\underline{\nu}_1}=
	\frac{1}{\mathpzc{q}},
	$$
	the inclusion $\overline{\mathcal{P}}\subset\underline{\mathcal{P}}$ happens
	if and only if $\mathpzc{q}=\frac{1}{m}$. In this case,
	we have $\overline{\nu}_1=m\underline{\nu}_1=(m+1)\nu_1$.
\end{rmk}

\begin{figure}[ht!]
	\centering
	\begin{psfrags}	
	\psfrag{a}{$\nu$}
	\psfrag{b}{$\underline{\nu}$}
	\psfrag{c}{$\overline{\nu}$}
	\psfrag{d}{\!\!\!\!\!\!\!\!$\underline{\nu}_{\mbox{\tiny 5}}\!\!=\!\overline{\nu}_{\mbox{\tiny 3}}$}
	\includegraphics[scale = 1.3]{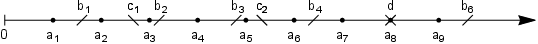}
	\caption{The set $\mathcal{P}$ and the values of $\nu_n$ for 
	$x_0=\frac{L}{8}$.}
	\label{line}
\end{psfrags}
\end{figure}

\noindent {\bf Example.} For $x_0=\frac{L}{8}$, we obtain distribution
sketched in Figure~\ref{line}.

\subsection{The amplitude of $\frac{\Psi_\nu}{\rho_\nu}$,
as a function of $\nu$, in the case that $x_0=0$}

The amplitude of the
oscillations in the case $x_0=0$ changes with the angular wave number.
Let us see how. For $x_0=0$, the expression for $\frac{\Psi_\nu}{\rho_\nu}$
simplifies to
\begin{equation}\label{chica}
	\frac{1}{\rho_\nu}\Psi_\nu(x)=
	\begin{cases}
		\frac{1}{\sqrt{1-\,\frac{\sin\left(\frac{L\nu}{2}\right)}{\frac{L\nu}{2}}}}
		\sqrt{\frac{2}{L}}
		\sin\left(\frac{\nu}{2}
		\left(\frac{L}{2}+x\right)\right)&\text{for}\ -\,\frac{L}{2}<x<0,\\
		\frac{1}{\sqrt{1-\,\frac{\sin\left(\frac{L\nu}{2}\right)}{\frac{L\nu}{2}}}}
		\sqrt{\frac{2}{L}}\sin\left(\frac{\nu}{2}
		\left(\frac{L}{2}-x\right)\right)&\text{for}\ 0<x<\frac{L}{2}.
	\end{cases}
\end{equation}
Let us set $\gamma=\frac{L\nu}{2}$
and $\Gamma(\gamma):=\frac{1}{\sqrt{1-\,\frac{\sin
			\left(\gamma\right)}{\gamma}}}$.
It is the factor $\Gamma$
that governs the amplitude of the wave functions.
Moreover,
$$\nu\in(\nu_{n-1},\nu_{n+1})\ \Rightarrow\ \gamma\in((n-1)\pi,(n+1)\pi).$$ 
Fix $n$ to be odd.
Taking into account the sign of $\sin \gamma$, we have
\begin{eqnarray*}
	\gamma\in((n-1)\pi,n\pi)&\Rightarrow&\Gamma(\gamma)>1,\\
	\gamma\in(n\pi,(n+1)\pi)&\Rightarrow&\Gamma(\gamma)<1.
\end{eqnarray*}
As we increase~$\nu$, 
the amplitude of the wave functions
oscillates according to the function $\Gamma$.
In Appendix~\ref{app} we take a closer look at the amplitude
of the oscillations.

\section{Zero eigenvalue}\label{zero}
When $E=0$, the solution of~\eqref{eq1} is piecewise linear:
\begin{equation}\label{psizero}
	\frac{\Psi_0}{\rho_0}(x) = \begin{cases}
		\frac{4\sqrt{3}}{\sqrt{L}(L^2-4x_0^2)}
		\left(\frac{L}{2}-x_0\right)\left(\frac{L}{2}+x\right), &\text{for}\ -\frac{L}{2}<x<x_0 \\
		\frac{4\sqrt{3}}{\sqrt{L}(L^2-4x_0^2)}
		\left(\frac{L}{2}+x_0\right)\left(\frac{L}{2}-x\right), & 
		\text{for}\ x_0<x<\frac{L}{2}.
	\end{cases} 
\end{equation}
This function has $L^2$ norm equal to $1$.
Equation~\eqref{eq4} becomes
$$
2g=-\,\frac{L}{\frac{L^2}{4}-x_0^2}=:2g_0=:2\frac{m\alpha_0}{\hbar^2}.
$$
Note that
$$
\lim_{\nu\searrow 0}f(\nu)
=2g_0.
$$
Moreover, as one might expect, we have
$$
\frac{
		\int_{x_0}^{\frac{L}{2}}\Psi_0^2(x)\,dx}
	{\int_{-\frac{L}{2}}^{x_0}\Psi_0^2(x)\,dx}=\mathpzc{q}.
$$
Furthermore,
$$
E_0(x):=\int_{-\,\frac{L}{2}}^{\frac{L}{2}}x\left(\frac{\Psi_0}{\rho_0}\right)^2(x)\,dx
=\frac{x_0}{2}.
$$

\section{Negative eigenvalues}\label{neg}
\begin{figure}[ht!]
	\begin{psfrags}
		\psfrag{n}{$\nu$}
		\psfrag{f}{\!\!\!\!\!$h(\nu)$}
		\centering
		\includegraphics[scale = .8]{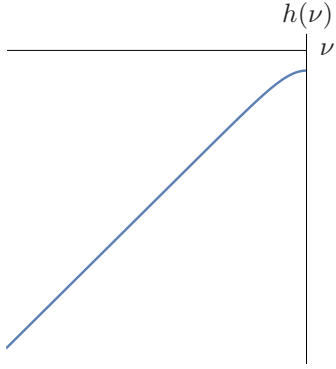}
		\caption{Sketch of the graph of $h$ for $x_0 = \frac{L}{8}$.}
		\label{fig2}
	\end{psfrags}
\end{figure}

We now turn to the case when $E<0$. 
Since $\sin(ix)=i\sinh(x)$, and the Schr\"{o}dinger equation
is linear, the function
\begin{equation}\label{psinegativo}
	\Psi_{i\nu}(x) = \begin{cases}
		\sinh\left(\frac{\nu }{2}\left(\frac{L}{2} - x_0\right)\right) \sinh\left(\frac{\nu}{2}\left( \frac{L}{2} + x\right)\right), &\text{for}\ -\frac{L}{2}<x<x_0 \\
		\sinh\left(\frac{\nu }{2}\left(\frac{L}{2} + x_0\right)\right) \sinh\left(\frac{\nu}{2}\left( \frac{L}{2} - x\right)\right), & 
		\text{for}\ x_0<x<\frac{L}{2},
	\end{cases} 
\end{equation}
satisfies~\eqref{eq1} with
$$
E=\frac{\hbar^2}{2m}\left(\frac{i\nu}{2}\right)^2=-\,
\frac{\hbar^2}{2m}\left(\frac{\nu}{2}\right)^2.
$$  
and
\begin{eqnarray*}
	2g &=&  h(\nu)\ :=\ f(i\nu)\\&=&
-\frac{\nu}{2}\frac{\sinh\left({ \frac{\nu}{2}L}\right)}{\sinh\left({\frac{\nu}{2}\left(\frac{L}{2} + x_0\right)}\right) \sinh\left({\frac{\nu}{2}\left(\frac{L}{2} - x_0\right)}\right)}\\
&=&-\,\frac{\nu}{2}\left(
\coth\left(\frac{\nu}{2}\left(\frac{L}{2}-x_0\right)\right)+
\coth\left(\frac{\nu}{2}\left(\frac{L}{2}+x_0\right)\right)
\right)
\end{eqnarray*}
(see \cite[(3)]{J}).
Note that all the functions involved in these formulas are even in~$\nu$.
So we may, and we will, take $\nu$ to be negative.
In this way, as $\nu$ runs through the negative real axis we obtain 
negative eigenvalues, and as $\nu$ runs through the positive real axis
we obtain the positive eigenvalues.
The last expression implies that $h'(\nu)>0$ for $\nu<0$ (and $h'(0)=0$),
because 
$$
\left(-x\coth x\right)'=\frac{x-\,\frac{1}{2}\sinh(2x)}{\sinh^2x}>0\ \text{for}\ x<0.
$$
Moreover, we have
$$
\lim_{\nu\nearrow 0}h(\nu)=2g_0\quad \text{and}\quad
\lim_{\nu\to-\infty}\frac{h(\nu)}{\nu}=1.
$$
The graph of $h$ for $x_0 = \frac{L}{8}$ is sketched in Figure~\ref{fig2}.
Choose $\rho_{i\nu}>0$ equal to the positive square root of
\begin{eqnarray*}
	&&	\int_{-\,\frac{L}{2}}^{\frac{L}{2}}\Psi_{i\nu}^2(x)\,dx\label{norm_negative}\\
	&&\frac{1}{2} \left(
	\left(\frac L2+x_0\right) \left(\frac{\sinh\left(\nu\left(\frac L2+x_0 \right)\right)}{\nu\left(\frac L2+ x_0 \right)}-1\right) \sinh ^2
	\left(\frac{\nu}{2}\left(\frac L2- x_0 \right)\right)
	\right.\nonumber\\
	&&\ \ \ \ \ \left.+
	\left(\frac L2- x_0\right) \left(\frac{\sinh\left(\nu\left(\frac L2- x_0 \right)\right)}{\nu\left(\frac L2- x_0 \right)}-1\right) \sinh ^2
	\left(\frac{\nu}{2}\left(\frac L2+ x_0 \right)\right)
	\right).\nonumber
\end{eqnarray*}
The function $\frac{\Psi_{i\nu}}{\rho_{i\nu}}$ is a solution
of the Schr\"{o}dinger equation with $L^2$ norm equal to~$1$.
We also have that
\begin{eqnarray*}
&&\lim_{\nu \to -\infty}\frac{
	\int_{x_0}^{\frac{L}{2}}\Psi_{i\nu}^2(x)\,dx}
{\int_{-\frac{L}{2}}^{x_0}\Psi_{i\nu}^2(x)\,dx}\\
&&\qquad=\lim_{\nu \to -\infty}
 \frac{\sinh\left(\nu\left(\frac L2- x_0 \right)\right)-\nu\left(\frac L2- x_0\right) }
{\sinh ^2
	\left(\frac{\nu}{2}\left(\frac L2- x_0 \right)\right)}\times\\
	&&\qquad\qquad\times\lim_{\nu \to -\infty}\frac{\sinh ^2
	\left(\frac{\nu}{2}\left(\frac L2+ x_0 \right)\right)}
	{\sinh\left(\nu\left(\frac L2+ x_0 \right)\right)-\nu\left(\frac L2+ x_0\right)}\\
&&\qquad=(-2)\times\left(-\,\frac{1}{2}\right)\ =\ 1.
\end{eqnarray*}
For each $\nu<0$, define the function $\underline{\Psi}_\nu:\left[-\,\frac{L}{2},\frac{L}{2}\right]\to\mathbb{R}$, by
$$
\underline{\Psi}_\nu(x):=
\begin{cases}
	\sqrt{-\,\frac{\nu}{2}}e^{-\,\frac{\nu}{2}(x-x_0)},&\text{for}\ -\,\frac{L}{2}\leq x\leq x_0,\\
	\sqrt{-\,\frac{\nu}{2}}e^{\frac{\nu}{2}(x-x_0)},&\text{for}\ x_0<x\leq\frac{L}{2}.
\end{cases}
$$
One can easily check that, for each $x\in\left(-\,\frac{L}{2},\frac{L}{2}\right)$,
we have
$$
\lim_{\nu\to-\infty}\frac{\,\,\frac{\Psi_{i\nu}}{\rho_{i\nu}}(x)\,\,}{\underline{\Psi}_\nu(x)}=1.
$$
For each small $\epsilon>0$, $\underline{\Psi}_\nu(x_0-\epsilon)$ and
 $\underline{\Psi}_\nu(x_0+\epsilon)$ converge to zero as $\nu\to-\infty$.
 As $\Psi_{i\nu}$ is increasing in $\left[-\,\frac{L}{2},x_0\right]$ and
 is decreasing in $\left[x_0,\frac{L}{2}\right]$, we conclude that
 $\frac{\Psi_{i\nu}}{\rho_{i\nu}}$ converges uniformly to zero
 as $\nu\to -\infty$, both in
 $\left[-\,\frac{L}{2},x_0\right]$ and
 in $\left[x_0,\frac{L}{2}\right]$.
 Therefore, as $\nu\to -\infty$, $\frac{\Psi_{i\nu}}{\rho_{i\nu}}$ converges to $\delta(x-x_0)$
 in the sense of the distributions.
This implies that
	$$
\lim_{\nu \to -\infty}
\int_{-\,\frac{L}{2}}^{\frac{L}{2}}x\left(\frac{\Psi_{i\nu}}{\rho_{i\nu}}\right)^2(x)\,dx=x_0,
$$
an equality that may also be checked by direct computation.
We see that as our angular wave number 
(which is $\frac{i\nu}{2}$ for negative $\nu$ in this section,
and is $\frac{\nu}{2}$ for positive $\nu$ in Section~\ref{pos})
moves along the negative imaginary axis, up to the origin,
and then moves along the positive real axis, towards~$+\infty$,
we obtain the full set of solutions $\Psi_\nu$.
To get other complex angular wave numbers
we should drop the restriction $\alpha\in\mathbb{R}$.
 For $i\nu$ imaginary, with $\nu$ negative, by abuse of notation, we write
$\Psi_\nu$ and $\rho_\nu$, instead of
$\Psi_{i\nu}$ and $\rho_{i\nu}$.

\section{Examples illustrating the behavior of the solutions in each of
the subintervals of $\mathbb{R}$ determined by $\mathcal{P}$}\label{exemplos}

Let us consider three examples.

\subsection{Case where $x_0=0$.} 
\begin{figure}[ht!]
	\centering
	\begin{psfrags}	
		\psfrag{b}{\footnotesize $x_0$}
		\includegraphics[scale = .8]{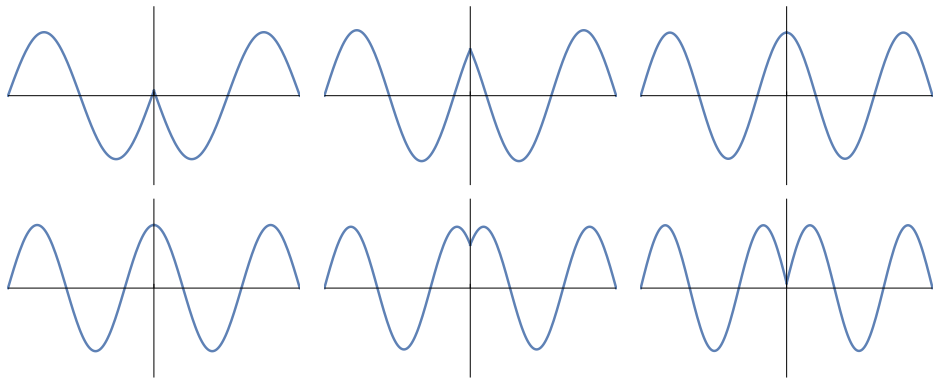}
		\caption{$x_0=0$. Sketch of the graph of $\frac{\Psi_\nu}{\rho_\nu}$ as $\nu$ varies in the subinterval 
			$(\nu_4,\nu_6)$}
		\label{x0}
	\end{psfrags}
\end{figure}

We want to see the
evolution of $\frac{\Psi_\nu}{\rho_\nu}$ as $\nu$ varies in the subinterval 
$(\nu_4,\nu_6)$
containing 
$\nu_5$, as in~\eqref{D}. This is the situation studied in~\cite{jviana11}.
See Figure~\ref{x0}.

{\em Across each set of six plots in this section, {\rm Section~\ref{exemplos}},
the first plot corresponds to a~$\nu$ slightly above the infimum of the 
subinterval, the second plot corresponds to a~$\nu$ which is 
equal to the average of $\nu_n$ and the infimum of the subinterval,
the third and forth plots correspond to the~$\nu$ $\nu_n$,
the fifth plot corresponds to a~$\nu$ which is equal to the average of
$\nu_n$ and the supremum of the subinterval,
and the sixth plot corresponds to a~$\nu$ slightly below the supremum of 
the subinterval.
Uniform scaling has been applied,
so that one can see how the amplitudes of the wave functions are changing with~$\nu$.}

In the case of Figure~\ref{x0}, it is hard to see
the change in amplitude, since it is of the order of~3\%.

\subsection{Case where $x_0=\frac{L}{8}$}

Having looked into the case $x_0=0$, we want to consider some other
$x_0=\lambda\frac{L}{2}$, where $\lambda$ has to belong to the interval $(0,1)$.
We want to choose $\lambda$ so that 
all
five possible types of subintervals, labeled~\eqref{A}, \eqref{B}, \eqref{C}, \eqref{E} and
\eqref{F}, arise. So, we are restricted to a rational $\lambda$ such that
we don't have
$\overline{\mathcal{P}}\subset\underline{\mathcal{P}}$. 
In the set
$\left\{\frac{1}{2}, \frac{1}{3}, \frac{2}{3}, \frac{1}{4}, \frac{3}{4}
\right\}$, $\lambda=\frac{1}{4}$ is the only choice. So we choose
$x_0=\frac{L}{8}$.

Fix a $\nu_n$, corresponding to a natural $n$. 
We want to see the
evolution of $\frac{\Psi_\nu}{\rho_\nu}$ as $\nu$ increases in the subinterval 
defined by $\mathcal{P}$ containing 
$\nu_n$, as $\alpha$ runs from $-\infty$ to $+\infty$. In the case that $\nu_n$ belongs to
$\mathcal{P}$, we just sketch $\Phi_{\nu_n}$;
applying Remark~\ref{q} to the present case, this case happens when 
$\nu_n$ belongs to~$\mathcal{K}$ and
$n$ is a multiple of $8$ (recall Figure~\ref{line}).

As $\nu\nearrow\nu_8$ and as $\nu\searrow\nu_8$, 
the amplitude of the oscillations on the right of~$x_0$ becomes 
$\frac{1}{\mathpzc{q}}=\frac{5}{3}$ times
the amplitude of the oscillations on the left of~$x_0$. 
The ratio of the probability of finding the particle
on the right of $x_0$ over the probability of finding the particle on the left of $x_0$ becomes $\frac{5}{3}$ (see~\eqref{prob}).
Compare the last plot in Figure~\ref{sete} and the first plot in Figure~\ref{nove}
to the plot of Figure~\ref{oito}.

Below are the evolutions of $\frac{\Psi_\nu}{\rho_\nu}$ for $\nu$ in each 
type of subinterval of $\mathbb{R}$. 

\begin{itemize}
	\item{$\nu_1\in(-\infty,\underline{\nu}_1)$.} 
	Figure~\ref{negative} illustrates the evolution for 
	$\nu$ in $(-\infty, 0)$. Figure~\ref{um} illustrates the evolution for 
	$\nu$ in $(0,\underline{\nu}_1)$, as in~\eqref{G}.
	
	\begin{figure}[ht!]
		\centering
		\begin{psfrags}	
			\psfrag{b}{\footnotesize $x_0$}
			\includegraphics[scale = .8]{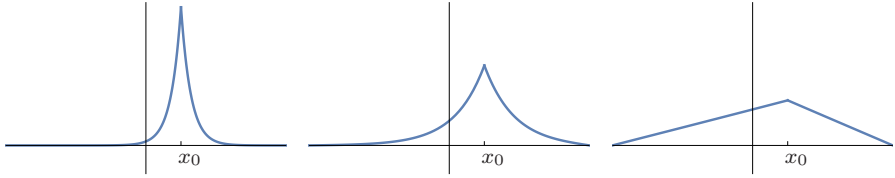}
			\caption{$x_0=\frac{L}{8}$. Sketch of the graph of $\frac{\Psi_\nu}{\rho_\nu}$ as $\nu$ varies in the subinterval 
				$(-\infty,0)$.
				The plot to the left corresponds to $\nu=-9\nu_1$,
				the middle plot corresponds to 
				$\nu=-3\nu_1$,
				and the plot on the right corresponds to
				$\nu=-\,\frac{1}{10}\nu_1$.}
			\label{negative}
		\end{psfrags}
	\end{figure}	

\begin{figure}[ht!]
	\centering
	\begin{psfrags}	
		\psfrag{b}{\footnotesize $x_0$}
		\includegraphics[scale = .8]{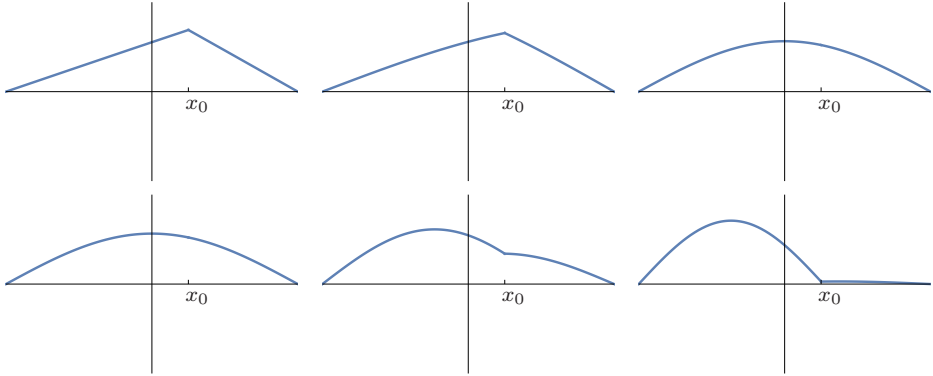}
		\caption{$x_0=\frac{L}{8}$. Sketch of the graph of $\frac{\Psi_\nu}{\rho_\nu}$ as $\nu$ varies in the subinterval 
			$(0,\underline{\nu}_1)$.}
		\label{um}
	\end{psfrags}
\end{figure}

\item{$\nu_2\in(\underline{\nu}_1,\overline{\nu}_1)$, $\nu_5\in(\underline{\nu}_3,\overline{\nu}_2)$, as in~\eqref{B}.} See Figure~\ref{dois} for the first case.

\begin{figure}[ht!]
	\centering
	\begin{psfrags}	
		\psfrag{b}{\footnotesize $x_0$}
		\includegraphics[scale = .8]{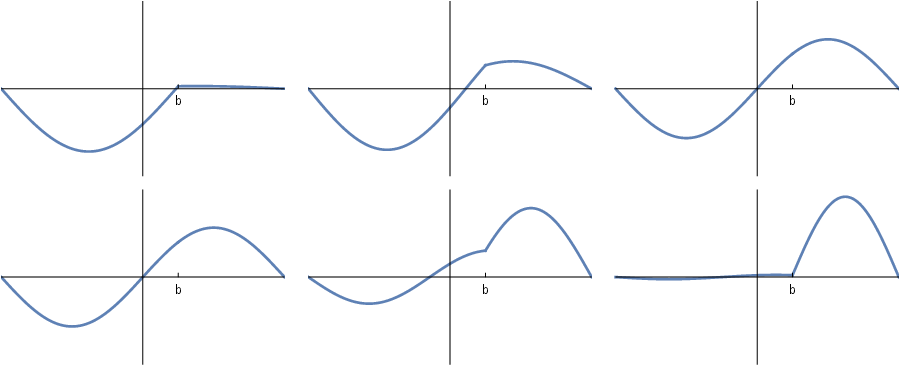}
		\caption{$x_0=\frac{L}{8}$. Sketch of the graph of $\frac{\Psi_\nu}{\rho_\nu}$ as $\nu$ varies in the subinterval 
			$(\underline{\nu}_1,\overline{\nu}_1)$.}
		\label{dois}
	\end{psfrags}
\end{figure}

\item{$\nu_3\in(\overline{\nu}_1,\underline{\nu}_2)$, $\nu_6\in(\overline{\nu}_2,\underline{\nu}_4)$, as in~\eqref{C}.}
See Figure~\ref{tres} for the first case.

\begin{figure}[ht!]
	\centering
	\begin{psfrags}	
		\psfrag{b}{\footnotesize $x_0$}
		\includegraphics[scale = .8]{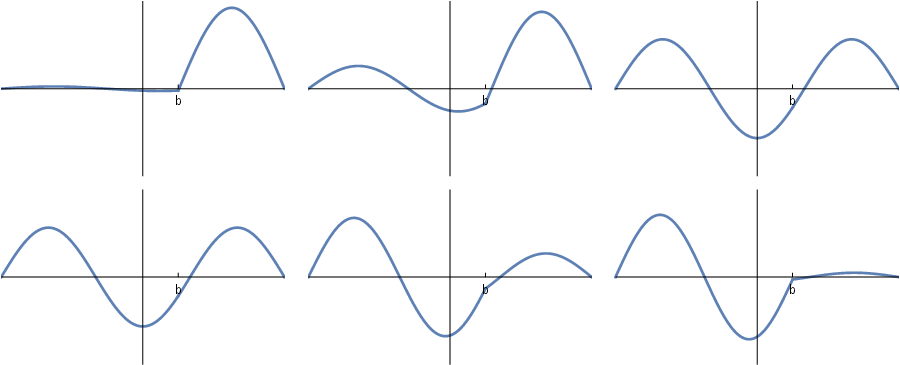}
		\caption{$x_0=\frac{L}{8}$. Sketch of the graph of $\frac{\Psi_\nu}{\rho_\nu}$ as $\nu$ varies in the subinterval 
			$(\overline{\nu}_1,\underline{\nu}_2)$.}
		\label{tres}
	\end{psfrags}
\end{figure}

\item{$\nu_4\in(\underline{\nu}_2,\underline{\nu}_3)$, as in~\eqref{A}.}
See Figure~\ref{quatro}.

\begin{figure}[ht!]
	\centering
	\begin{psfrags}	
		\psfrag{b}{\footnotesize $x_0$}
		\includegraphics[scale = .8]{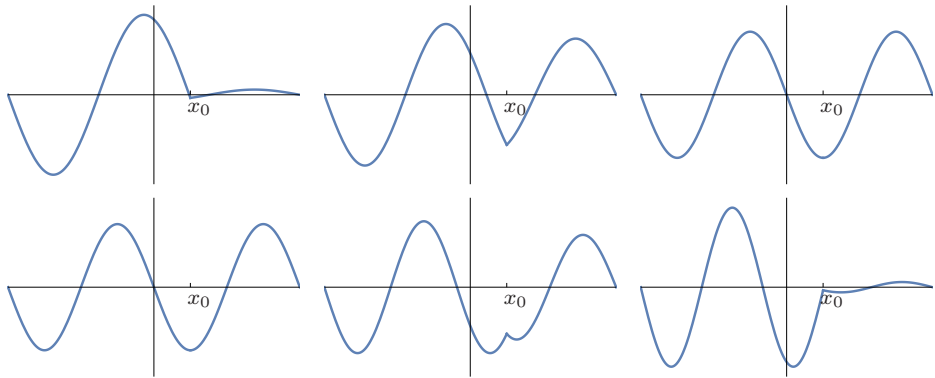}
		\caption{$x_0=\frac{L}{8}$. Sketch of the graph of $\frac{\Psi_\nu}{\rho_\nu}$ as $\nu$ varies in the subinterval 
			$(\underline{\nu}_2,\underline{\nu}_3)$.}
		\label{quatro}
	\end{psfrags}
\end{figure}

\item{$\nu_7\in(\underline{\nu}_4,{\nu}_8)$, as in~\eqref{F}.}
See Figure~\ref{sete}.

\begin{figure}[ht!]
	\centering
	\begin{psfrags}	
		\psfrag{b}{\footnotesize $x_0$}
		\includegraphics[scale = .8]{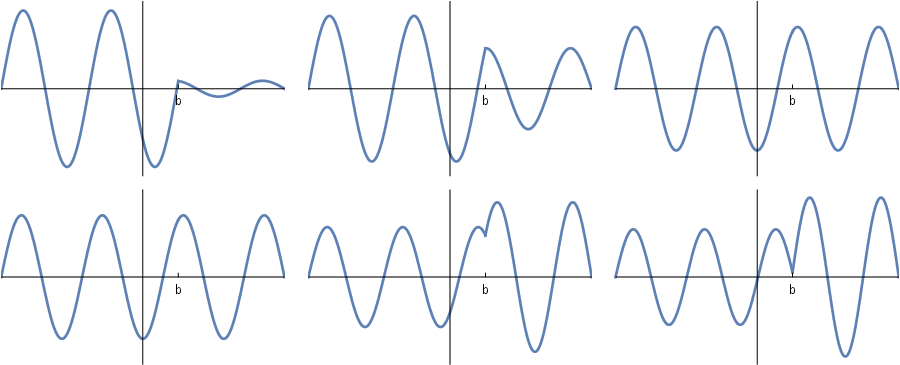}
		\caption{$x_0=\frac{L}{8}$. Sketch of the graph of $\frac{\Psi_\nu}{\rho_\nu}$ as $\nu$ varies in the subinterval 
			$(\underline{\nu}_4,{\nu}_8)$.}
		\label{sete}
	\end{psfrags}
\end{figure}

\item{$\nu_8=\underline{\nu}_5=\overline{\nu}_3$, as in~\eqref{Z}.}
See Figure~\ref{oito}.

\begin{figure}[ht!]
	\centering
	\begin{psfrags}	
		\psfrag{b}{\footnotesize $x_0$}
		\includegraphics[scale = .8]{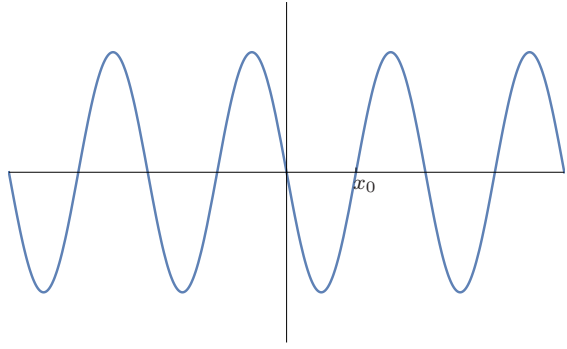}
		\caption{$x_0=\frac{L}{8}$. Sketch of the graph of $\Phi_{\nu_8}$.}
		\label{oito}
	\end{psfrags}
\end{figure}

\item{$\nu_9\in({\nu}_8,\underline{\nu}_6)$, as in~\eqref{E}.}
See Figure~\ref{nove}.

\begin{figure}[ht!]
	\centering
	\begin{psfrags}	
		\psfrag{b}{\footnotesize $x_0$}
		\includegraphics[scale = .8]{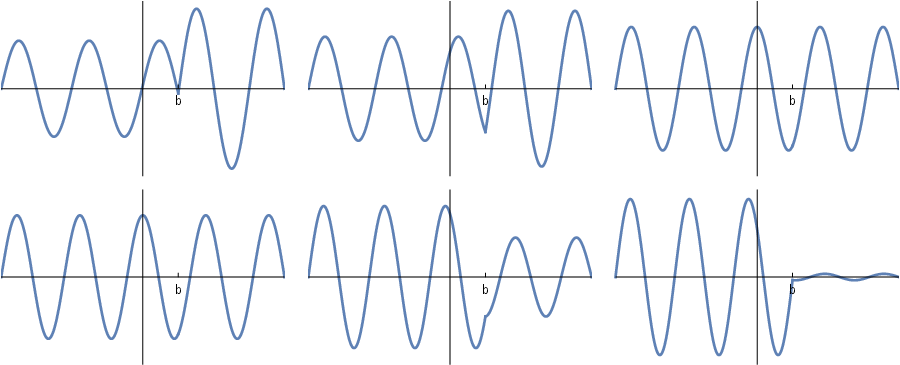}
		\caption{$x_0=\frac{L}{8}$. Sketch of the graph of $\frac{\Psi_\nu}{\rho_\nu}$ as $\nu$ varies in the subinterval 
			$({\nu}_8,\underline{\nu}_6)$.}
		\label{nove}
	\end{psfrags}
\end{figure}

\end{itemize}

\subsection{Case where $x_0=\frac{L}{\sqrt{63}}$}\label{surprise}

In this subsection we give an example of a 
situation where $\underline{\nu}_k<\underline{\nu}_{k+1}$
are two consecutive values of~$\underline{\mathcal{P}}$, with no value
of~$\overline{\mathcal{P}}$ between them,
such that as we increase~$\nu$ 
from~$\underline{\nu}_{k}$ to~$\underline{\nu}_{k+1}$, we find values
of $\nu$ for which $\Psi_\nu$ describes a particle with a high probability of being on the right of~$x_0$.
 
\begin{figure}[ht!]
	\begin{psfrags}
		\psfrag{b}{$x_0$}
		\psfrag{x}{$x$}
		\psfrag{y}{\!\!\!\!\!\!$\frac{\Psi_{\check{\nu}}}{\rho_{\check{\nu}}}(x)$}
		\centering
		\includegraphics[scale = .8]{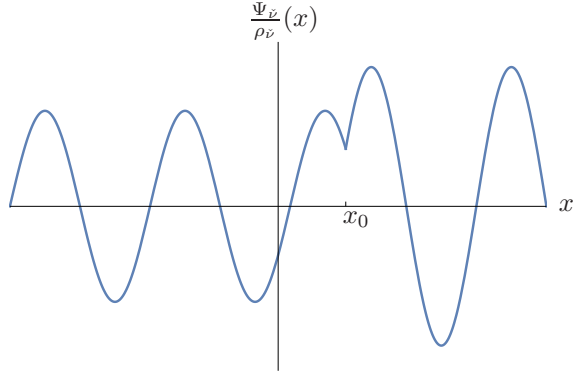}
		\caption{Sketch of the graph of $\frac{\Psi_{\check{\nu}}}{\rho_{\check{\nu}}}$ for $x_0 = \frac{L}{\sqrt{63}}$ and $\check{\nu}=\nu_7+\frac{2}{3}(\underline{\nu}_5-\nu_7)$.
			This profile is similar to the one on the bottom right of Figure~\ref{sete}.}
		\label{right}
	\end{psfrags}
\end{figure}

We take $x_0=\frac{L}{\sqrt{63}}$. Note that
$\frac{1}{\sqrt{63}}\approx 0.126\approx 0.125=\frac{1}{8}$.
We have
$\nu_7\in(\underline{\nu}_4,\underline{\nu}_5)$, and we have
$\underline{\nu}_5\approx 0.998\, \nu_8$ (remember that
$\nu_n$ does not depend on $x_0$).
Let us choose 
$\nu\in(\underline{\nu}_4,\underline{\nu}_5)$ close, but not too 
close to
$\underline{\nu}_5$. For example
 $\nu=\check{\nu}:=\nu_7+\frac{2}{3}(\underline{\nu}_5-\nu_7)$ will do.
The reason for this choice will become clear in Figure~\ref{media63}.
Then, in spite of $\underline{\Upsilon}_{\underline{\nu}_4}$
and $\underline{\Upsilon}_{\underline{\nu}_5}$ being zero on the right of $x_0$,
the function
$\frac{\Psi_{\check{\nu}}}{\rho_{\check{\nu}}}$ has a profile
similar to the one of the function $\Upsilon_{\nu_8}$ that corresponds
to the choice $x_0=\frac{L}{8}$,
which is sketched in the bottom right of Figure~\ref{sete}.
It corresponds to a particle
that has a higher probability of being on the right of $\frac{L}{\sqrt{63}}$
than of being on the left of $\frac{L}{\sqrt{63}}$
(the ratio of the two probabilities is approximately equal to
$1.28$;
for the function in the bottom right of Figure~\ref{sete}, 
the ratio of the probability of finding a particle
on the right of $\frac{L}{8}$ over the probability of finding a particle
on the left of $\frac{L}{8}$ is approximately equal to
$1.66$).
See Figure~\ref{right}.
One important difference between the solutions sketched in
the bottom right of Figure~\ref{sete} and in Figure~\ref{right}
consists in the fact that in the first case the value of $\alpha$ is
very large, while in the second case the value of $\alpha$ is just equal to
$\frac{\hbar^2}{2m}f(\check{\nu})\approx 13.5\,\frac{\hbar^2}{2m}$.

\section{The behavior of the ratio of the probability of finding the particle on the right of $x_0$ over the probability of finding the particle on the left of $x_0$, as a function of $\nu$, and the expectation value of
	the position of the particle, as a function of~$\nu$.}\label{pr}

We define a function $r:\mathbb{R}\setminus\mathcal{P}\to\mathbb{R}$ by
\begin{equation}\label{probability}
	r(\nu):=\frac{
		\int_{x_0}^{\frac{L}{2}}\Psi_\nu^2(x)\,dx}
	{\int_{-\frac{L}{2}}^{x_0}\Psi_\nu^2(x)\,dx},
\end{equation}
and we extend it by continuity to $\mathcal{K}$. Since we still call the
extension $r$, the function $r$ is defined in 
$\mathcal{K}\cup\mathbb{R}\setminus\mathcal{P}$. For $\hat{\nu}\in\mathcal{K}$, 
recall, from Remark~\ref{nice}, that we have
$$
r(\hat{\nu})=\frac{
	\int_{x_0}^{\frac{L}{2}}\Upsilon_{\hat{\nu}}^2(x)\,dx}
{\int_{-\frac{L}{2}}^{x_0}\Upsilon_{\hat{\nu}}^2(x)\,dx}
=\frac{1}{\mathpzc{q}},\
\text{for}\ \hat{\nu}=\underline{\nu}_k=\overline{\nu}_l=\nu_{k+l}.
$$
This value is different from 
$$
\frac{
	\int_{x_0}^{\frac{L}{2}}\Phi_{\nu_{k+l}}^2(x)\,dx}
{\int_{-\frac{L}{2}}^{x_0}\Phi_{\nu_{k+l}}^2(x)\,dx}
=\mathpzc{q},$$
as long as $x_0$ is not equal to zero. We mention that in the case that
$x_0$ is zero, we have symmetry about zero, and so $r\equiv 1$.

We have seen that we always have $r(0)=\mathpzc{q}$
and that function $r$ always tends to~$1$ as~$\nu$ tends to~$-\infty$.
If $\underline{\nu}\in
\underline{\mathcal{P}}\setminus\mathcal{K}$, then
$\lim_{\nu\to\underline{\nu}}r(\nu)=0$, while if $\overline{\nu}\in
\overline{\mathcal{P}}\setminus\mathcal{K}$, then
$\lim_{\nu\to\overline{\nu}}r(\nu)=+\infty$.
We also point out that, if $\nu_n\not\in\mathcal{K}$, then
$$
r(\nu_n)=\mathpzc{q}\,\frac{1-\,\frac{\sin\left(n\pi\left(
		1-\,\frac{2x_0}{L}\right)\right)}{n\pi\left(
		1-\,\frac{2x_0}{L}\right)}}{1-\,\frac{\sin\left(n\pi\left(
		1+\frac{2x_0}{L}\right)\right)}{n\pi\left(
		1+\frac{2x_0}{L}\right)}}\ \longrightarrow\ \mathpzc{q},\ \ \text{as}\
	n\to+\infty.
$$

The expectation value of
the position of the particle with wave function~$\frac{\Psi_\nu}{\rho_\nu}$ is given by
$$
E_\nu(x):=
\int_{-\,\frac{L}{2}}^{\frac{L}{2}}x
\left(\frac{\Psi_\nu}{\rho_\nu}\right)^2(x)\,dx.
$$
Let us calculate the expectation value for $\Upsilon_{\hat{\nu}}$, for $\hat{\nu}\in\mathcal{K}$.
Recall that if $\hat{\nu}$ belongs to $\mathcal{K}$, the there exists a natural number
$n$ such that $\hat{\nu}=\nu_n$.
Taking into account that
\begin{eqnarray*}
&&	\int x\sin^2\left(\frac{\nu}{2}\left(\frac{L}{2}\pm x\right)\right)\,dx\\
&&\qquad\qquad=
	\int\frac{x}{2}\left(1-\cos\left(\nu\left(\frac{L}{2}\pm x\right)\right)\right)\,dx\\
	&&\qquad\qquad=\frac{1}{2\nu^2}\left(
	\frac{\nu^2x^2}{2}\mp \nu x\sin\left(\nu\left(\frac{L}{2}\pm x\right)\right)
	-\cos\left(\nu\left(\frac{L}{2}\pm x\right)\right)
	\right),
\end{eqnarray*}
in the case that $\nu$ equals some $\nu_n$
we get
\begin{eqnarray*}
	\int_{-\,\frac{L}{2}}^{x_0} x\sin^2\left(\frac{\nu}{2}\left(\frac{L}{2}+ x\right)\right)\,dx&=&
	\frac{1}{4}\left(
	x_0^2-\,\frac{L^2}{4}
	\right),\\
	\int^{\frac{L}{2}}_{x_0} x\sin^2\left(\frac{\nu}{2}\left(\frac{L}{2}- x\right)\right)\,dx&=&
	\frac{1}{4}\left(\frac{L^2}{4}-
	x_0^2
	\right).
\end{eqnarray*}
The previous two integrals add up to $0$, as they should, because the
expectation value of a particle described by a function $\Phi_{\nu_n}$ is $0$.
So, in the case that $\nu$ equals some $\nu_n$, we have
\begin{eqnarray*}
&&	\frac{\frac{L}{2}-x_0}{\frac{L}{2}+x_0}
	\int_{-\,\frac{L}{2}}^{x_0} x\sin^2\left(\frac{\nu}{2}\left(\frac{L}{2}- x\right)\right)\,dx+
	\frac{\frac{L}{2}+x_0}{\frac{L}{2}-x_0}
	\int^{\frac{L}{2}}_{x_0} x\sin^2\left(\frac{\nu}{2}\left(\frac{L}{2}- x\right)\right)\,dx
\\ &&\qquad\qquad =
	\frac{Lx_0}{2}.
\end{eqnarray*}
Using the definition of $\Upsilon_{\hat{\nu}}$, we obtain
$$
\int_{-\,\frac{L}{2}}^{\frac{L}{2}}x\Upsilon_{\hat{\nu}}^2(x)\,dx=x_0.
$$

We define $\overline{l}$ and $\overline{r}$ to be the midpoints 
of $\left(-\,\frac{L}{2},
x_0\right)$ and $\left(x_0,\frac{L}{2}\right)$, respectively:
$$
\overline{l}:=\frac{1}{2}\left(-\,\frac{L}{2}+x_0\right),\qquad
\overline{r}:=\frac{1}{2}\left(x_0+\frac{L}{2}\right).
$$

\subsection{Two examples in the case where $x_0$ is a rational multiple of $L$}

\begin{figure}[ht!]
	\centering
	\begin{psfrags}	
		\psfrag{a}{\footnotesize $\mathpzc{q}$}
\psfrag{c}{\footnotesize $1$}
\psfrag{b}{\footnotesize $\frac{1}{\mathpzc{q}}$}
\psfrag{n}{\footnotesize $\nu$}
\psfrag{r}{\footnotesize \!\!\!\!$r(\nu)$}
		\includegraphics[scale = 1.2]{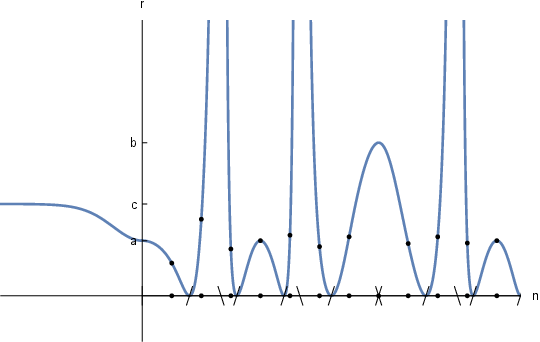}
		\caption{Sketch of the graph of $r$ for $x_0=\frac{L}{8}$.}
		\label{ratio}
	\end{psfrags}
\end{figure}

\begin{figure}[ht!]
	\centering
	\begin{psfrags}	
		\psfrag{n}{$\nu$}
\psfrag{o}{$\!\nu_8$}
\psfrag{x}{\!\!\!\!\!$x_0$}
\psfrag{l}{\!\!$\overline{l}$}
\psfrag{r}{\!\!$\overline{r}$}
\psfrag{L}{\!\!\!$\frac{L}{2}$}
\psfrag{K}{\!\!\!\!\!\!\!\!$-\,\frac{L}{2}$}
\psfrag{m}{\!\!\!\!\!$E_\nu(x)$}
		\includegraphics[scale = 1.2]{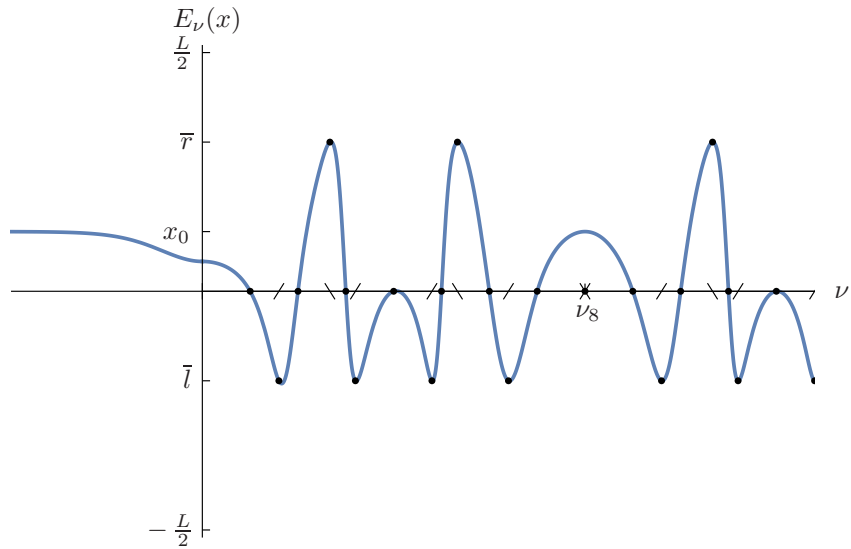}
		\caption{Sketch of the graph of the expectation value, $E_\nu(x)$, of
			the position of a particle with 
			wave function $\frac{\Psi_\nu}{\rho_\nu}$
			when $x_0=\frac{L}{8}$. 
			A particle with wave function $\Upsilon_{\nu_8}$
			has $E_{\nu_8}(x)=x_0$.}
		\label{media8}
	\end{psfrags}
\end{figure}

In Figure~\ref{ratio} we sketch the graph of $r$ and in Figure~\ref{media8}
we sketch the graph of $\nu\mapsto E_{\nu}(x)$ when 
$x_0=\frac{L}{8}$. 

\begin{figure}[ht!]
	\centering
	\begin{psfrags}	
		\psfrag{n}{$\nu$}
		\psfrag{y}{\!$\check{\nu}$}
		\psfrag{o}{$\!\nu_8$}
		\psfrag{x}{\!\!\!\!\!$x_0$}
		\psfrag{l}{\!\!$\overline{l}$}
		\psfrag{r}{\!\!$\overline{r}$}
		\psfrag{L}{\!\!\!$\frac{L}{2}$}
		\psfrag{K}{\!\!\!\!\!\!\!\!$-\,\frac{L}{2}$}
		\psfrag{m}{\!\!\!\!\!$E_\nu(x)$}
		\includegraphics[scale = 1.2]{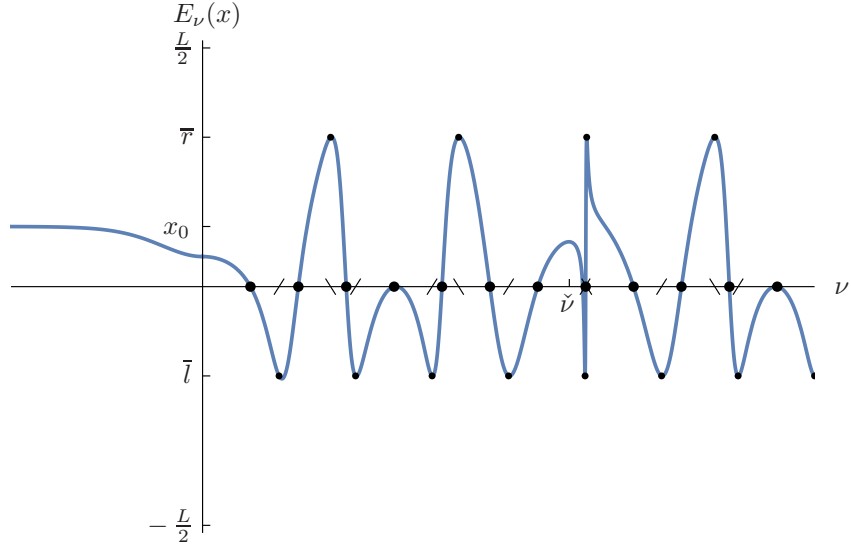}
		\caption{Sketch of the graph of the expectation value, $E_\nu(x)$, of
			the position of a particle with 
			wave function $\frac{\Psi_\nu}{\rho_\nu}$
			when $x_0=\frac{L}{\sqrt{63}}$.}
		\label{media63}
	\end{psfrags}
\end{figure}

If we change $x_0=\frac{L}{8}$ to $x_0=\frac{L}{\sqrt{63}}$,
then the plot of Figure~\ref{media8} changes to the plot of
Figure~\ref{media63}. Note that between $\underline{\nu}_4$ 
and $\underline{\nu}_5$ we have marked
$\check{\nu}=\nu_7+\frac{2}{3}(\underline{\nu}_5-\nu_7)$,
whose wave function is sketched in Figure~\ref{right}.

\begin{figure}[ht!]
	\centering
	\begin{psfrags}	
		\psfrag{a}{\footnotesize $\mathpzc{q}$}
		\psfrag{c}{\footnotesize $1$}
		\psfrag{b}{\footnotesize $\frac{1}{\mathpzc{q}}$}
		\psfrag{n}{\footnotesize $\nu$}
		\psfrag{r}{\footnotesize \!\!\!\!$r(\nu)$}
		\includegraphics[scale = 1.2]{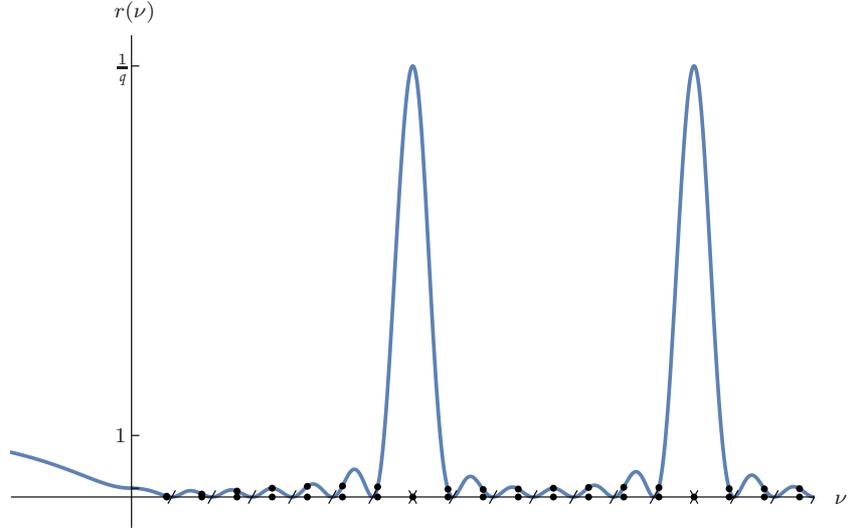}
		\caption{Sketch of the graph of $r$ for $x_0=\frac{3L}{8}$.}
		\label{ratio45}
	\end{psfrags}
\end{figure}

\begin{figure}[ht!]
	\centering
	\begin{psfrags}	
		\psfrag{n}{$\nu$}
		\psfrag{o}{$\!\nu_8$}
		\psfrag{x}{\!\!\!\!\!$x_0$}
		\psfrag{l}{\!\!$\overline{l}$}
		\psfrag{r}{\!\!$\overline{r}$}
		\psfrag{L}{\!\!\!$\frac{L}{2}$}
		\psfrag{K}{\!\!\!\!\!\!\!\!$-\,\frac{L}{2}$}
		\psfrag{m}{\!\!\!\!\!$E_\nu(x)$}
		\includegraphics[scale = 1.2]{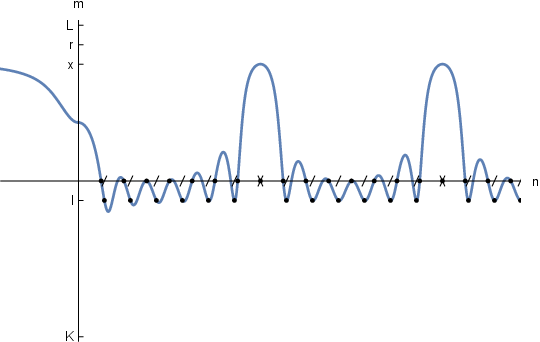}
		\caption{Sketch of the graph of the expectation value, $E_\nu(x)$, of
			the position of a particle with 
			wave function $\frac{\Psi_\nu}{\rho_\nu}$
			when $x_0=\frac{3L}{8}$.}
		\label{media34}
	\end{psfrags}
\end{figure}

In the next example, we consider the value of $m$ of
Remark~\ref{m} equal to~$7$, and
in Figure~\ref{ratio45} we sketch the graph of $r$ when 
$x_0=\frac{3L}{8}$. 
As $\overline{\nu}_1=7\underline{\nu}_1$, $r$ never
blows up. 
The function $r$ goes to zero at $\underline{\nu}_k$, 
for $k$ a non natural multiple of~$7$.
When $\underline{\nu}_k$ coincides with some $\overline{\nu}_l$,
in this case for $k=7l$, the ratio~$r$ is~$7$.
In Figure~\ref{media34}
we sketch the graph of $\nu\mapsto E_{\nu}(x)$ when $x_0=\frac{3L}{8}$.

\subsection{An example in the case where $x_0$ is an irrational multiple of $L$}

\begin{figure}[ht!]
	\centering
	\begin{psfrags}	
		\psfrag{a}{\footnotesize $\mathpzc{q}$}
		\psfrag{c}{\footnotesize $1$}
		\psfrag{b}{\footnotesize $?$}
		\psfrag{n}{\footnotesize $\nu$}
		\psfrag{r}{\footnotesize \!\!\!\!$r(\nu)$}
		\includegraphics[scale = 1.2]{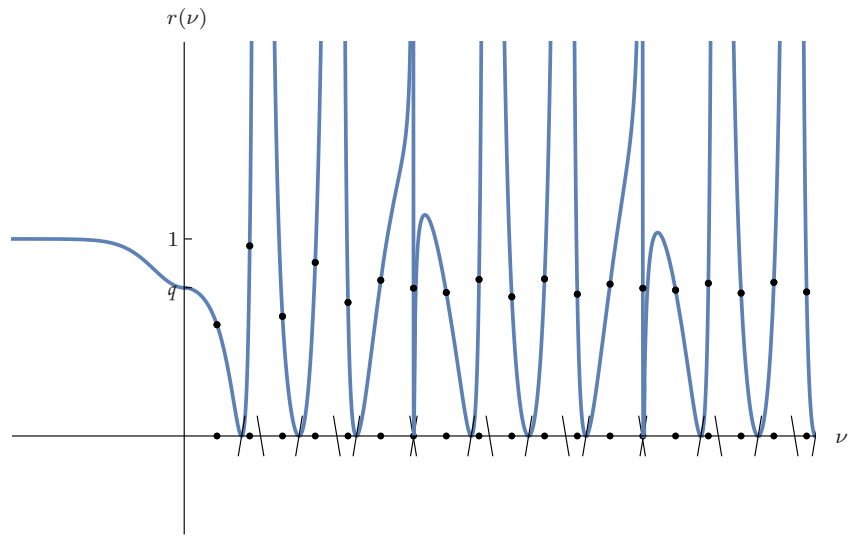}
		\caption{Sketch of the graph of $r$ for $x_0=\frac{L}{10\sqrt{2}}$.}
		\label{ratio50}
	\end{psfrags}
\end{figure}

In Figure~\ref{ratio50} we sketch the graph of $r$ when 
$x_0$ is not a rational multiple of $L$. Specifically, we choose
$x_0=\frac{L}{10\sqrt{2}}$.
A small $x_0$ brings $\underline{\nu}_1$ and $\overline{\nu}_1$
closer, and thus leads to a richer structure in an interval with a fixed 
length, which we depict in our plot. 
In this figure, it looks like
$\overline{\nu}_3$, $\nu_7$ and $\underline{\nu}_4$ coincide, but in fact
$\overline{\nu}_3<\nu_7<\underline{\nu}_4$.
So, we also have that
$\overline{\nu}_6<\nu_{14}<\underline{\nu}_8$.

\begin{figure}[ht!]
	\centering
	\begin{psfrags}	
		\psfrag{a}{\footnotesize $\mathpzc{q}$}
		\psfrag{c}{\footnotesize $1$}
		\psfrag{b}{\footnotesize $\frac{1}{\mathpzc{q}}$}
		\psfrag{n}{\footnotesize $\nu$}
		\psfrag{r}{\footnotesize \!\!\!\!$r(\nu)$}
		\includegraphics[scale = 1.2]{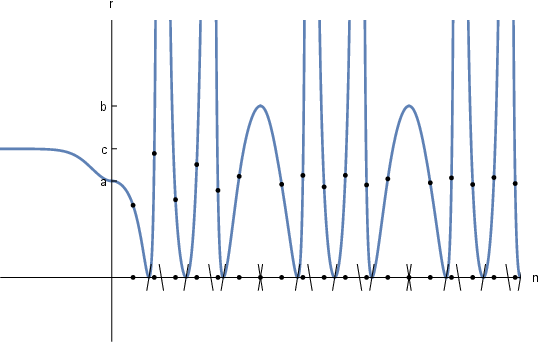}
		\caption{Sketch of the graph of $r$ for $x_0=\frac{L}{14}$.}
		\label{ratio14}
	\end{psfrags}
\end{figure}

Noteworthy is the way $r$ blows up as $\nu\nearrow\overline{\nu}_3$
(and as $\nu\nearrow\overline{\nu}_6$).
The graph seems to wiggle, it has an inflection point.
Let us see what is happening.
If, instead of choosing $x_0=\frac{L}{10\sqrt{2}}$, we choose
$x_0=\frac{L}{14}$, then we obtain $\overline{\nu}_3=\nu_{7}=\underline{\nu}_4
=\frac{14\pi}{L}$.
In Figure~\ref{ratio14} we sketch the graph of $r$ when $x_0=\frac{L}{14}$.
The values of $\frac{1}{10\sqrt{2}}$ and $\frac{1}{14}$ are 
approximately equal to $0.0707$ and $0.0714$, respectively, so they 
match to the first two decimal places. 
So the inflection points in the graph of $r$ for $x_0=\frac{L}{10\sqrt{2}}$ 
happen because if we take a sequence $(x_0)_m\nearrow\frac{L}{14}$, as $m\to\infty$,
and we choose a $\nu$ close to $\frac{14\pi}{L}$,
different from $\frac{14\pi}{L}$, then
$\left(\underline{\nu}_4\right)_{(x_0)_m}\searrow
\left(\underline{\nu}_4\right)_{\frac{L}{14}}=
\frac{14\pi}{L}$,
$\left(\overline{\nu}_3\right)_{(x_0)_m}\nearrow
\left(\overline{\nu}_3\right)_{\frac{L}{14}}=
\frac{14\pi}{L}$, and
$r_{(x_0)_m}(\nu)$ will converge to $r_{\frac{L}{14}}(\nu)$, as $m\to\infty$.

If, on the other hand, we decided to approximate $\frac{1}{10\sqrt{2}}$
by a sequence of rational numbers, $\left(\frac{p_m}{q_m}\right)$,
then the sequence $(q_m)$ would have tend to infinity,
so that the values such that $\underline{\nu}_k=\overline{\nu}_l$
(which depend on $m$),
would also tend to $\infty$ (see Remark~\ref{q}).

\subsection{The case where $x_0$ is small}

As $x_0$ decreases to $0$, $\underline{\nu}_k$ increases to $\nu_{2k}$
and $\overline{\nu}_l$ decreases to $\nu_{2l}$. Hence, 
$$
x_0\searrow 0\ \Rightarrow\ \underline{\nu}_m\nearrow \nu_{2m}\ \text{and}\
\overline{\nu}_m\searrow\nu_{2m}.
$$
Let $\epsilon>0$ be small. 
There exists $\delta>0$ such that if $0<x_0<\delta$, then
$$\underline{\nu}_{m-1}<\overline{\nu}_{m-1}<\nu_{2m-2}+\epsilon$$ and
$$\overline{\nu}_m>\underline{\nu}_m>\nu_{2m}-\epsilon.$$
Thus, 
for $0<x_0<\delta$, 
the interval $[\nu_{2m-2}+\epsilon,\nu_{2m}-\epsilon]$ does not contain any point
of $\mathcal{P}$.
This implies that we are in the case of Subsection~\ref{third}, and 
so~\eqref{psi} and~\eqref{norm} hold. Using this last equality,
and taking into account that
$$
\lim_{x_0\to 0}\frac{\sin^2\left(\frac{\nu}{2}\left(\frac{L}{2}+x_0\right)\right)}
{\sin^2\left(\frac{\nu}{2}\left(\frac{L}{2}-x_0\right)\right)}=1
$$
and
$$
\lim_{x_0\to 0}\frac{1-\,\frac{\sin\left(\nu\left(\frac{L}{2}-x_0\right)\right)}{\nu\left(\frac{L}{2}-x_0\right)}}
{1-\,\frac{\sin\left(\nu\left(\frac{L}{2}+x_0\right)\right)}{\nu\left(\frac{L}{2}+x_0\right)}}=1,
$$
we obtain
\begin{rmk}\label{rose}
	For each natural number $m$ and each small $\epsilon>0$, the
	restriction of the ratio $r$ to the interval
	$[\nu_{2m-2}+\epsilon,\nu_{2m}-\epsilon]$ converges uniformly to
	$1$, as $\epsilon$ converges to $0$.
\end{rmk}
Since $\epsilon$ is arbitrary, the restriction of the ratio $r$
to the open interval $(\nu_{2m-2},\nu_{2m})$ goes pointwise to $1$.

\begin{figure}[ht!]
	\centering
	\begin{psfrags}	
		\psfrag{a}{\footnotesize $\mathpzc{q}$}
		\psfrag{c}{\footnotesize $1$}
		\psfrag{b}{\footnotesize $\frac{1}{\mathpzc{q}}$}
		\psfrag{n}{\footnotesize $\nu$}
		\psfrag{r}{\footnotesize \!\!\!\!$r(\nu)$}
		\includegraphics[scale = 1.2]{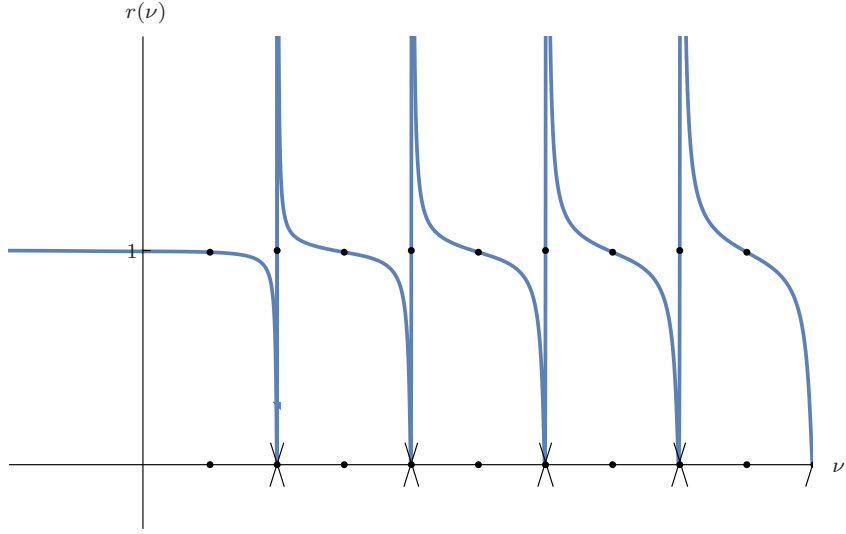}
		\caption{Sketch of the graph of $r$ for $x_0=\frac{L}{1000}$.}
		\label{muitopequeno}
	\end{psfrags}
\end{figure}

\begin{figure}[ht!]
	\centering
	\begin{psfrags}	
		\psfrag{n}{$\nu$}
		\psfrag{o}{$\!\nu_8$}
		\psfrag{x}{\!\!\!\!\!$x_0$}
		\psfrag{l}{\!\!$\overline{l}$}
		\psfrag{r}{\!\!$\overline{r}$}
		\psfrag{L}{\!\!\!$\frac{L}{2}$}
		\psfrag{K}{\!\!\!\!\!\!\!\!$-\,\frac{L}{2}$}
		\psfrag{m}{\!\!\!\!\!$E_\nu(x)$}
		\includegraphics[scale = 1.2]{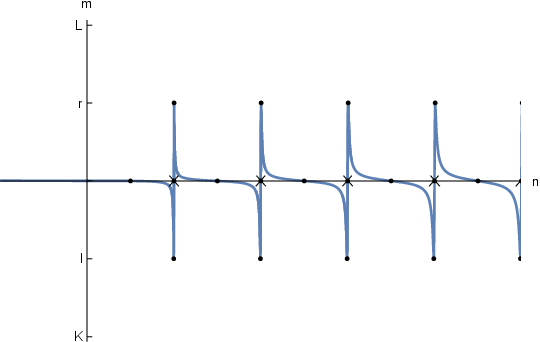}
		\caption{Sketch of the graph of the expectation value, $E_\nu(x)$, of
			the position of a particle with 
			wave function $\frac{\Psi_\nu}{\rho_\nu}$
			when $x_0=\frac{L}{1000}$.}
		\label{media0}
	\end{psfrags}
\end{figure}

To illustrate Remark~\ref{rose}, in Figure~\ref{muitopequeno} we sketch the graph of 
$r$ when $x_0=\frac{L}{1000}$.
In Figure~\ref{media0}
we sketch the graph of $\nu\mapsto E_{\nu}(x)$ when $x_0=\frac{L}{1000}$.

\subsection{The case where $x_0$ is close to $\frac{L}{2}$}

Finally, we look into the case where $x_0$ is close to $\frac{L}{2}$.
For $0<\nu\leq\frac{\pi}{\frac{L}{2}-x_0}=\frac{\overline{\nu}_1}{2}$, we have
$$
\frac{2}{3}\leq
\frac{1-\,\frac{\sin\left(\nu\left(\frac{L}{2}-x_0\right)\right)}{\nu\left(\frac{L}{2}-x_0\right)}}{\sin^2\left(\frac{\nu}{2}\left(\frac{L}{2}-x_0\right)\right)}\leq 1,
$$
and
\begin{equation}\label{aux}
\lim_{x_0\to\frac{L}{2}}\frac{1-\,\frac{\sin\left(\nu\left(\frac{L}{2}-x_0\right)\right)}{\nu\left(\frac{L}{2}-x_0\right)}}{\sin^2\left(\frac{\nu}{2}\left(\frac{L}{2}-x_0\right)\right)}=\frac{2}{3}.
\end{equation}
So, using~\eqref{norm}, for $0<\nu\leq\frac{\overline{\nu}_1}{2}$, we have 
$$
\frac{2\mathpzc{q}}{3}\,
\frac{\sin^2\left(\frac{\nu}{2}\left(\frac{L}{2}+x_0\right)\right)}
{1-\,\frac
	{\sin\left(\nu\left(\frac{L}{2}+x_0\right)\right)}{\nu\left(\frac{L}{2}+x_0\right)}}
\leq
r(\nu)
\leq
\mathpzc{q}
\frac{\sin^2\left(\frac{\nu}{2}\left(\frac{L}{2}+x_0\right)\right)}
{1-\,\frac
	{\sin\left(\nu\left(\frac{L}{2}+x_0\right)\right)}{\nu\left(\frac{L}{2}+x_0\right)}}
	\leq\frac{3\mathpzc{q}}{2}.
$$
Moreover, again from~\eqref{norm} and from~\eqref{aux}, for $0<\nu\leq\frac{\overline{\nu}_1}{2}$, we have
$$
\lim_{x_0\to\frac{L}{2}}\frac{r(\nu)}{\mathpzc{q}}=\frac{2}{3}\,
\frac{\sin^2\left(\frac{\nu L}{2}\right)}
{1-\,\frac
	{\sin\left(\nu L\right)}{\nu L}}\leq 1.
$$
Since $\overline{\nu}_1$ goes to $+\infty$ as $x_0$ goes to $\frac{L}{2}$,
and since $\lim_{x_0\nearrow\frac{L}{2}}\mathpzc{q}=0$,
this establishes
\begin{rmk}\label{gardenia}
 Given $M>0$, as $x_0\nearrow\frac{L}{2}$,
the restriction of the ratio $r$ to the interval $[0,M]$
goes uniformly to zero.
\end{rmk}
However, according to Remark~\ref{m}, we have, for each natural number $m$,
$$
x_0=\frac{m-1}{m+1}\frac{L}{2}\ \Rightarrow\ r\left((m+1)\frac{2\pi}{L}\right)=m,
$$
and so the pointwise convergence of $r$ to zero is not uniform in $\mathbb{R}^+_0$.

\begin{figure}[ht!]
	\centering
	\begin{psfrags}	
		\psfrag{a}{\footnotesize $\mathpzc{q}$}
		\psfrag{c}{\footnotesize $1$}
		\psfrag{b}{\footnotesize $\frac{1}{\mathpzc{q}}$}
		\psfrag{n}{\footnotesize $\nu$}
		\psfrag{r}{\footnotesize \!\!\!\!$r(\nu)$}
		\includegraphics[scale = 1.2]{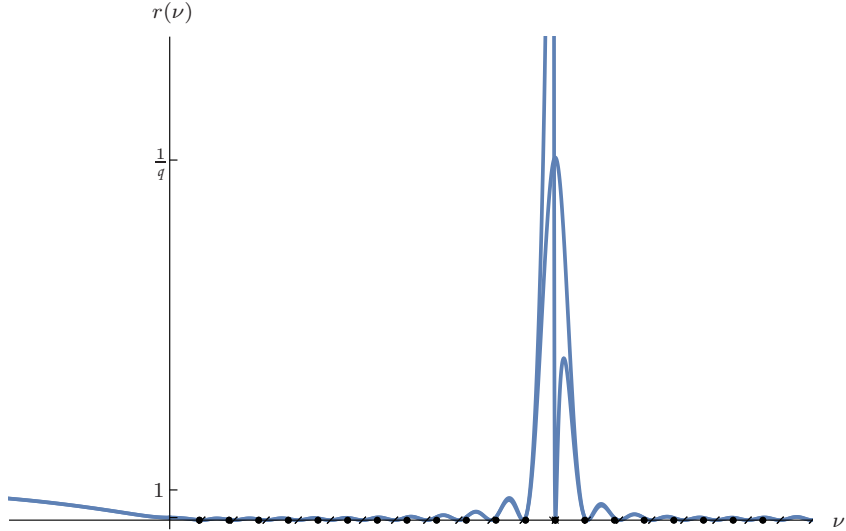}
		\caption{Superimposed sketches of the graphs of $r$ for $x_0=\sqrt{\frac{5}{7}}\frac{L}{2}$
			and 
			$x_0=\frac{11}{13}\frac{L}{2}$.}
		\label{superimposed}
	\end{psfrags}
\end{figure}

We illustrate Remark~\ref{gardenia} with a sketch of the graph of
the ratio $r$ for two relatively large values of $x_0$, one an
irrational multiple of $L$, and the other one
of the form
$x_0=\frac{m-1}{m+1}\frac{L}{2}$, with $m=12$.
We choose the two values of $x_0$ close together.
In Figure~\ref{superimposed}, we superimpose sketches of the graphs of $r$
for $x_0=\sqrt{\frac{5}{7}}\frac{L}{2}$ and for
$x_0=\frac{11}{13}\frac{L}{2}$. Approximate values of
$\sqrt{\frac{5}{7}}$ and $\frac{11}{13}$ are
$0.8451$
and
$0.8461$,
respectively.
Given that when $x_0=\frac{11}{13}\frac{L}{2}$ we have
$\overline{\mathcal{P}}\subset\underline{\mathcal{P}}$,
in this case,
qualitatively, we obtain a plot like the one of Figure~\ref{ratio45},
but with a smaller $\mathpzc{q}$. 
 We did not choose an even bigger values of $x_0$
because this would move the peak even further to the right
and make it even taller.

\appendix

\section{The Fourier series of $\frac{\Psi_\nu}{\rho_\nu}$}\label{serie}

In this appendix, we calculate the Fourier series of $\frac{\Psi_\nu}{\rho_\nu}$.
Consider first an interval $\mathcal{I}$ contained in $\mathbb{R}^+\setminus\mathcal{P}$
and let $n$ be the natural number such that $\nu_n\in\mathcal{I}$.
Clearly, for $\nu=\nu_n$,
$$
\frac{\Psi_{\nu_n}}{\rho_{\nu_n}}(x)=\Phi_{\nu_n}(x).
$$
Let $\nu\in\mathcal{I}$, with $\nu$ different from $\nu_n$.
Let us calculate the Fourier series coefficients such that
$$
\frac{\Psi_\nu}{\rho_\nu}(x)=\sum_{m=1}^\infty a_m\Phi_{\nu_m}(x).
$$
This series converges uniformly since $\frac{\Psi_\nu}{\rho_\nu}$ is piecewise $C^1$.
Testing~\eqref{eq1} with $\Phi_{\nu_m}$, we obtain
\begin{eqnarray*}
&&-\frac{\hbar^{2}}{2 m} \int_{-\frac{L}{2}}^{\frac{L}{2}}
\frac{\Psi_\nu}{\rho_\nu}(x)\frac{d^2\Phi_{\nu_m}}{d{x}^2}(x)\,dx + \alpha \int_{-\frac{L}{2}}^{\frac{L}{2}}\delta(x - x_0)\frac{\Psi_\nu}{\rho_\nu}(x) \Phi_{\nu_m}(x)\,dx\\
&&\qquad\qquad\qquad\qquad\qquad\qquad\qquad\qquad\qquad\qquad = E_{{\nu}}\int_{-\frac{L}{2}}^{\frac{L}{2}} \frac{\Psi_\nu}{\rho_\nu}(x)\Phi_{\nu_m}(x)\,dx.
\end{eqnarray*}
Since
$-\Phi_{\nu_m}''=\frac{\pi^2m^2}{L^2}
\Phi_{\nu_m}$, 
$E_{{\nu}}=\frac{\hbar^2}{2m}\left(\frac{{\nu}}{2}\right)^2$ and 
$\alpha=\frac{\hbar^2}{2m}f(\nu)$, we obtain
$$ 
	\frac{\pi^2m^2}{L^2}a_m+
	f(\nu)\frac{\Psi_\nu(x_0)}{\rho_\nu}
	\Phi_{\nu_m}(x_0)=\frac{\nu^2}{4}a_m.
$$ 
As
$$
f(\nu)\Psi_\nu(x_0)=-\,\frac{\nu}{2}\sin\left(\frac{\nu}{2}L\right)
(-1)^{\left\lfloor\frac{\nu}{2}\left(\frac{L}{2}+x_0\right)\frac{1}{\pi}
	\right\rfloor}
$$
and there is no value of $m$ for which $\frac{\pi^2m^2}{L^2}=\frac{\nu^2}{4}$,
solving for $a_m$, we get
\begin{eqnarray*}
	a_m&=&(-1)^{\left\lfloor\frac{\nu}{2}\left(\frac{L}{2}+x_0\right)\frac{1}{\pi}
		\right\rfloor}\frac{\nu}{2\rho_\nu}\sin\left(\frac{\nu}{2}L\right)\frac{\Phi_{\nu_m}(x_0)}
	{\frac{\pi^2m^2}{L^2}-\frac{\nu^2}{4}},
\end{eqnarray*}
for all $m\in\mathbb{N}$. 
This shows that
\begin{equation}\label{Fourier_4}
\frac{\Psi_\nu}{\rho_\nu}(x)=\tilde{c}_{x_0,\nu}
	\sum_{m=1}^\infty
	\frac{\Phi_{\nu_m}(x_0)}{\frac{\pi^2m^2}{L^2}-\frac{\nu^2}{4}}\Phi_{\nu_m}(x)
\end{equation}
for $\nu\in\mathcal{I}\setminus\{\nu_n\}$,
with
$$
\tilde{c}_{x_0,\nu}:=(-1)^{\left\lfloor\frac{\nu}{2}\left(\frac{L}{2}+x_0\right)\frac{1}{\pi}
	\right\rfloor}\frac{\nu}{2\rho_\nu}\sin\left(\frac{\nu}{2}L\right).
$$
Given that
$
\lim_{\nu\to\nu_n}\tilde{c}_{x_0,\nu}=0
$,
if $m\neq n$ then 
$
\lim_{\nu\to\nu_n}a_m=0,
$
as had to be the case
(because the map $\nu\mapsto\frac{\Psi_\nu}{\rho_\nu}$
is continuous in $C^0$).
Seeing that
$$
\lim_{\nu\to\nu_n}\frac{\sin\left(\frac{\nu}{2}L\right)}{\frac{\pi n}{L}-\,\frac{\nu}{2}}=-L(-1)^n,
$$
$$
\lim_{\nu\to\nu_n}\frac{\nu}{\frac{\pi n}{L}+\frac{\nu}{2}}=1,
$$
$$
\lim_{\nu\to\nu_n}\frac{\Phi_{\nu_n}(x_0)}{\rho_\nu}=\frac{2}{L}(-1)^{n+1}
(-1)^{\left\lfloor\frac{\nu}{2}\left(\frac{L}{2}+x_0\right)\frac{1}{\pi}
	\right\rfloor},
$$
we confirm that
$$
\lim_{\nu\to\nu_n}
\frac{\,\Phi_{\nu_n}(x_0)
	\tilde{c}_{x_0,\nu}\,}{\frac{\pi^2n^2}{L^2}-\frac{\nu^2}{4}}=1.
$$
The Fourier series
\eqref{Fourier}, \eqref{Fourier_2} and \eqref{Fourier_3} are
limit cases of~\eqref{Fourier_4}. 

One can also check that
$$
\lim_{\nu\searrow 0}a_m=
\lim_{\nu\searrow 0}
\frac{\,\Phi_{\nu_m}(x_0)\tilde{c}_{x_0,\nu}\,}{\frac{\pi^2m^2}{L^2}-\frac{\nu^2}{4}}=
\frac{4\sqrt{3}}{L^2-4x_0^2}
\frac{\sqrt{2}L^2}{\pi^2m^2}
\sin\left(\frac{m\pi}{L}\left(\frac{L}{2}-x_0\right)\right),
$$
and, obviously, these are the Fourier coefficients of the function $\frac{\Psi_0}{\rho_0}$
in~\eqref{psizero}:
$$ 
\frac{\Psi_0}{\rho_0}(x)=\tilde{c}_{x_0,0}
\sum_{m=1}^\infty
\frac{\,\Phi_{\nu_m}(x_0)\,}{m^2}\Phi_{\nu_m}(x),
$$ 
with
$$
\tilde{c}_{x_0,0}:=\frac{4\sqrt{3}L^{\frac{5}{2}}}{\pi^2(L^2-4x_0^2)}.
$$

Consider now $\nu<0$. We obtain
$$ 
	\frac{\Psi_\nu}{\rho_\nu}(x)=\tilde{c}_{x_0,\nu}
	\sum_{m=1}^\infty
	\frac{\Phi_{\nu_m}(x_0)}{\frac{\pi^2m^2}{L^2}+\frac{\nu^2}{4}}\Phi_{\nu_m}(x),
$$ 
with
$$
\tilde{c}_{x_0,\nu}:=\frac{\nu}{2\rho_\nu}\sinh\left(\frac{\nu}{2}L\right).
$$
This $\rho_\nu$ for $\nu$ negative is defined in Section~\ref{neg}.

\section{The amplitude of $\frac{\Psi_\nu}{\rho_\nu}$, as a function of $\nu$, in the case that $x_0=0$} \label{app}

Let us set $\gamma_n=n\pi$.
Note that when $\nu=\nu_n$, then $\gamma=\frac{L\nu}{2}=\gamma_n$.
Call $\overline{\gamma}_n$ the value of $\gamma\in((n-1)\pi,n\pi)$,
where $\gamma\mapsto\frac{\sin\gamma}{\gamma}$ attains its maximum,
and $\underline{\gamma_n}$ the value of $\gamma\in(n\pi,(n+1)\pi)$,
where $\gamma\mapsto\frac{\sin\gamma}{\gamma}$ attains its minimum.
In the next remark
we show that,
when $x_0=0$,
 the oscillations in the amplitude
of the wave function
 behave like
$\frac{1}{2n\pi}=\frac{1}{2\gamma_n}=\frac{1}{L\nu_n}$.
Precisely, we have
\begin{rmk}\label{swing}
	When $x_0=0$, the amplitude of the $\frac{\Psi_\nu}{\rho_\nu}$ satisfies
	\begin{eqnarray*}
		\lim_{\stackrel{\mbox{\tiny $n\to\infty$}}{\mbox{\tiny $n$ odd}}}
		\frac{\Gamma(\overline{\gamma}_n)-1}{\frac{1}{2\overline{\gamma}_n}}&=&1,\\
		\lim_{\stackrel{\mbox{\tiny $n\to\infty$}}{\mbox{\tiny $n$ odd}}}
		\frac{1-\Gamma(\underline{\gamma_n})}{\frac{1}{2\underline{\gamma_n}}}&=&1.
	\end{eqnarray*}	
	Moreover,
	\underline{for $n$ odd} (recall~\eqref{odd}), greater than or equal to $3$, we have that
	\begin{eqnarray*}
		\overline{\Gamma}_n:=
		\max_{\gamma\in((n-1)\pi,n\pi)}
		\Gamma(\gamma)&\in&
		\left(1+\frac{1}{2n\pi}+\frac{1}{3n^2\pi^2},\
		1+\frac{1}{2(n-1)\pi}+\frac{1}{2(n-1)^2\pi^2}\right),\\
		\underline{\Gamma}_n:=
		\min_{\gamma\in(n\pi,(n+1)\pi)}
		\Gamma(\gamma)&\in&
		\left(1-\frac{1}{2n\pi}+\frac{1}{3n^2\pi^2},\ 
		1-\frac{1}{2(n+1)\pi}+\frac{1}{2(n+1)^2\pi^2}\right).
	\end{eqnarray*}
\end{rmk}
\begin{proof}
	The function $\gamma\mapsto\frac{\sin\gamma}{\gamma}$ has
	its critical points when $\gamma=\tan\gamma$.
	This implies $\sin\gamma=\pm\frac{\gamma}{\sqrt{1+\gamma^2}}$
	and $\cos\gamma=\pm\frac{1}{\sqrt{1+\gamma^2}}$.
	Thus
	\begin{eqnarray*}
		\overline{\Gamma}_n&=&
		\frac{1}{\sqrt{1-\,\frac{1}{\sqrt{1+
						\overline{\gamma}_n^2}}}},\\
		\underline{\Gamma}_n&=&
		\frac{1}{\sqrt{1+\frac{1}{\sqrt{1+\underline{\gamma_n}^2}}}}.
	\end{eqnarray*}
	Expanding in Taylor series, we have
	\begin{eqnarray*}
		\overline{\Gamma}(\gamma)&:=&\frac{1}{\sqrt{1-\,\frac{1}{\sqrt{1+
						{\gamma}^2}}}}\ =\ 
		\frac{\sqrt[4]{1+\frac{1}{\gamma^2}}}
		{\sqrt{\sqrt{1+\frac{1}{\gamma^2}}-\,
				\frac{1}{\gamma}}}
		\\
		&=&1+\frac{1}{2\gamma}+\frac{3}{8\gamma^2}+
		O\left(\frac{1}{\gamma^3}\right),\\
		\underline{\Gamma}(\gamma)&:=&\frac{1}{\sqrt{1+\,\frac{1}{\sqrt{1+
						{\gamma}^2}}}}\ =\ 
		\frac{\sqrt[4]{1+\frac{1}{\gamma^2}}}
		{\sqrt{\sqrt{1+\frac{1}{\gamma^2}}+,
				\frac{1}{\gamma}}}\\
		&=&1-\frac{1}{2\gamma}+\frac{3}{8\gamma^2}+
		O\left(\frac{1}{\gamma^3}\right).
	\end{eqnarray*}
	Obviously, $\frac{1}{3}<\frac{3}{8}<\frac{1}{2}$.
	One can show that, for $\gamma\geq\pi$,
	$$
	\begin{array}{rcccl}
		1+\frac{1}{2\gamma}+\frac{1}{3\gamma^2}
		&\leq&\overline{\Gamma}(\gamma)	
		&\leq&1+\frac{1}{2\gamma}+\frac{1}{2\gamma^2},\\
		1-\,\frac{1}{2\gamma}+\frac{1}{2\gamma^2}
		&\leq&\underline{\Gamma}(\gamma)
		&\leq&1-\,\frac{1}{2\gamma}+\frac{1}{3\gamma^2}.
	\end{array}
	$$
	Hence, as 
	the functions
	$\gamma\mapsto 1-\frac{1}{2\gamma}+\frac{1}{2\gamma^2}$ and 
	$\gamma\mapsto 1-\,\frac{1}{2\gamma}+\frac{1}{3\gamma^2}$ are increasing 
	(for $\gamma>2$),
	and
	\begin{equation}\label{fair}
		\begin{array}{rcccl}
			\overline{\gamma}_n&\in&((n-1)\pi,n\pi)&=:&
			(\check{\overline{\gamma}}_n,
			\hat{\overline{\gamma}}_n),\\
			\underline{\gamma_n}&\in&(n\pi,(n+1)\pi)&=:&
			(\underline{\check{\gamma}_n},
			\underline{\hat{\gamma}_n}),
		\end{array}
	\end{equation}
	the remark is proved:
	\begin{eqnarray*}
		\overline{\Gamma}_n&\in&\left(1+\frac{1}{2\hat{\overline{\gamma}}_n}
		+\frac{1}{3\hat{\overline{\gamma}}_n^2},
		1+\frac{1}{2\check{\overline{\gamma}}_n}+\frac{1}{2\check{\overline{\gamma}}_n^2}\right),\\
		\underline{\Gamma}_n&\in&\left(1-\,\frac{1}{2
			\underline{\check{\gamma}_n}}
		+\,\frac{1}{3
			\underline{\check{\gamma}_n^2}}
		,
		1-\,\frac{1}{2\underline{\hat{\gamma}_n}}+
		\frac{1}{2\underline{\hat{\gamma}_n^2}}\right).
	\end{eqnarray*}
\end{proof}
The estimates in the previous remark can be improved if, instead of~\eqref{fair},
we obtain better estimates for $\overline{\gamma}_n$ and
$\underline{\gamma_n}$, the roots of $\gamma=\tan\gamma$ in
the intervals $((n-1)\pi,n\pi)$ and $(n\pi,(n+1)\pi)$, respectively.
One can show that
\begin{equation}\label{good}
	\begin{array}{rcl}
		\overline{\gamma}_n&\in&\left(\left(n-\frac{1}{2}\right)\pi-\,
		\frac{1}{3\left(n-\,\frac{1}{2}\right)},\left(n-\frac{1}{2}\right)\pi
		-\,
		\frac{1}{4\left(n-\frac{1}{2}\right)}\right),\\
		\underline{\gamma_n}&\in&\left(\left(n+\frac{1}{2}\right)\pi-\,
		\frac{1}{3\left(n+\frac{1}{2}\right)},\left(n+\frac{1}{2}\right)\pi-\,
		\frac{1}{4\left(n+\frac{1}{2}\right)}\right),
	\end{array}
\end{equation}
for $n\geq 2$. 

To provide insight into the precision of the estimates of Remark~\ref{swing},
let us mention that, for $n=3$, they reduce to
\begin{eqnarray*}
	\overline{\Gamma}_3&\approx&1.071\in(1.056,1.092),
	\\
	\underline{\Gamma}_3&\approx&0.957\in(0.950,0.963),
\end{eqnarray*}
while, for $n=33$, they reduce to
\begin{eqnarray*}
	\overline{\Gamma}_{33}&\approx&1.00493\in(1.00485,1.00502),
	\\
	\underline{\Gamma}_{33}&\approx&0.99528\in(0.99521,0.99536).
\end{eqnarray*}
The error is of the order $10^{-2}$ for $n=3$, and is of the order 
$10^{-4}$ for $n=33$.
Using~\eqref{good}, instead of~\eqref{fair},
	one obtains 
	\begin{eqnarray*}
		\overline{\Gamma}_3&\approx&1.0711\in(1.0700,1.0731),
		\\
		\underline{\Gamma}_3&\approx&0.9572\in(0.9569,0.9584),
	\end{eqnarray*}
	and
	\begin{eqnarray*}
		\overline{\Gamma}_{33}&\approx&1.004933\in(1.004929,1.004946),
		\\
		\underline{\Gamma}_{33}&\approx&0.995282\in(0.995278,0.995294).
	\end{eqnarray*}
	The error is of order~$10^{-3}$ for $n=3$, and is of order~$10^{-5}$ for
	$n=33$.

Consider now the case $n=1$. In the interval $(0,\pi)$, the
maximum of the wave function occurs at $x=0$, with
$$
\frac{1}{\rho_\nu}\Psi_\nu(0)=\sqrt{\frac{2}{L}}\sin\left(\frac{\gamma}{2}
\right)
\Gamma(\gamma).
$$
The function $\gamma\mapsto\sin\left(\frac{\gamma}{2}
\right)
\Gamma(\gamma)$ is strictly decreasing in the interval $(0,\pi)$.
Its value at $\pi$ is equal to $1$.
We have
$$
\overline{\Gamma}_1:=
\sup_{\gamma\in(0,\pi]}\left(\sin\left(\frac{\gamma}{2}
\right)
\Gamma(\gamma)\right)=
\lim_{\gamma\searrow 0}\left(\sin\left(\frac{\gamma}{2}
\right)
\Gamma(\gamma)\right)=
\sqrt{\frac{3}{2}}\approx 1.224.$$ 
In the interval $(\pi,2\pi)$,
the function $\Gamma$ is strictly below $1$,
tending to $1$ at the endpoints of the interval, and
with a minimum which is slightly above $0.9$:
$$\underline{\Gamma}_1=\min_{\gamma\in(\pi,2\pi)}\Gamma(\gamma)\approx 0.906.$$

\bibliographystyle{plain}
\bibliography{sample}

\end{document}